\documentclass[fleqn,usenatbib]{mnras}
\pdfoutput=1
\usepackage{hyperref}
\usepackage{amsmath}
\usepackage{multirow}
\usepackage{graphicx}
\usepackage[authoryear]{natbib}
\usepackage{amssymb,txfonts}
\providecommand{\tabularnewline}{\\}

\newcommand{\s}{\emph{Swift}}
\newcommand{\f}{\emph{Fermi}}

\begin{document}

\title{Modelling the luminosity function of long Gamma Ray Bursts using \s\, and \f\,}

\author[D. Paul]{Debdutta Paul\thanks{debdutta.paul@tifr.res.in}\\Tata Institute of Fundamental Research, India}

\maketitle

\global\long\def\AS{AstroSat}
\global\long\def\Ep{E_{p}}
\global\long\def\T{T_{90}}
\global\long\def\red{\chi_{{\rm red}}^{2}}

\begin{abstract}
I have used a sample of long Gamma Ray Bursts (GRBs) common to both \s\, and \f\, to re-derive the parameters of the Yonetoku correlation. This allowed me to self-consistently estimate pseudo redshifts of all the bursts with unknown redshifts. This is the first time such a large sample of GRBs from these two instruments are used, both individually and in conjunction, to model the long GRB luminosity function. The GRB formation rate is modelled as the product of the cosmic star formation rate and a GRB formation efficiency for a given stellar mass. An exponential cut-off powerlaw luminosity function fits the data reasonably well, with $\nu = 0.6$ and $ L_b = 5.4 \times10^{52} \, \rm{erg.s^{-1}},$ and does not require a cosmological evolution. In the case of a broken powerlaw, it is required to incorporate a sharp evolution of the break given by $L_{b}\sim0.3\times10^{52}\left(1+z\right)^{2.90} \, \rm{erg.s^{-1}},$ and the GRB formation efficiency (degenerate up to a beaming factor of GRBs) decreases with redshift as $\propto\left(1+z\right)^{-0.80}.$ However it is not possible to distinguish between the two models. The derived models are then used as templates to predict the distribution of GRBs detectable by CZTI on board $\AS$, as a function of redshift and luminosity. This demonstrates that via a quick localization and redshift measurement of even a few CZTI GRBs, $\AS$ will help in improving the statistics of GRBs both typical and peculiar.
\end{abstract}


\section{Introduction}

For any detector of gamma ray bursts (GRBs), an interesting estimable quantity is the number of GRBs observed, as a function of measurable parameters. This depends on instrumental parameters like duration-of-operation, $T$ and field-of-view $\Delta \Omega$, as well as the observed GRB production-rate and the distribution over intrinsic properties of GRBs. Following the formalism outlined in \citet{Tan_et_al.-2013-ApJL}, let us assume that the rate of GRBs beamed towards an observer on earth from an infinitesimal co-moving volume $dV,$ is given by $\overset{.}{R}\left(z\right)\frac{dV}{1+z},$ where $z$ denotes the redshift, and the factor $(1+z)^{-1}$ takes care of the cosmological time dilation.

Most generally, the number of GRBs detected by the instrument in the luminosity ($L$) range $L_{1}$ to $L_{2}$ and redshift range $z_{1}$ to $z_{2}$ is given by,

\begin{equation}
N(L_{1},L_{2};z_{1},z_{2})=T\,\dfrac{\Delta\Omega}{4\pi}\,\intop_{max[L_1,\, L_c]}^{L_2}\Phi_z(L)dL\,\intop_{z_{1}}^{z_{2}}\dfrac{\overset{.}{R}(z)}{1+z}dV,\label{eq:definition_of_phi}
\end{equation}

\noindent where $L_{c}$ denotes a lower-cutoff in the intrinsic luminosity of GRBs (see Section \ref{sec:The-estimated-luminosities}). The function $\Phi_z(L)$ is formally called the `luminosity function' (henceforth LF), having the units of $\rm{ ( erg.s^{-1} )^{-1} }, $ the subscript refering to an implicit dependence on the redshift. In view of the fact that GRBs are end-products of massive stars in galaxies, the GRB formation rate $\overset{.}{R}\left(z\right)$ can be written as
\begin{equation}
\overset{.}{R}\left(z\right)=f_{B}C\,\overset{.}{\rho_{\star}},\label{eq:R_dot}
\end{equation}

\noindent where $\overset{.}{\rho_{\star}}$ gives the cosmic star-formation rate (CSFR) in ${\rm M_{\odot}Gpc^{-3}yr^{-1}},$ $C$ gives the efficiency of GRB production given a certain stellar mass, in units of ${\rm M_{\odot}^{-1}},$ and $f_{B}$ encodes the beaming effect of the relativistic jets producing the burst. All of these terms may be functions of the redshift.

The dependence of the detected number distribution of a certain class of astrophysical object on its luminosity function, is quite general. It has been extensively studied in the context of galaxies, galaxy clusters (see \citet{Galaxy_cluster_LF-2017-MNRAS} and references therein), white dwarfs (see \citet{White_dwarf_LF-2016-review} for a recent review), quasars (see \citet{Quasar_LF_in_UV-2017-MNRAS} and references therein), high mass Xray binaries (see \citet{High_mass_XRB_LF-2017-MNRAS} and references therein) etc. The methodologies depend on the observational window available for the study of the particular objects of interest (e.g. while \citet{Galaxy_LF_using_WISE-2017-AJ}, \citet{Galaxy_LF_in_Kband-2017-MNRAS}, \citet{Galaxy_nearby_LF-2017-ApJ} etc. use the infrared bands to calculate the absolute magnitude of galaxies, \citet{Galaxy_LF_in_Bband-2017-A&A} use the optical B-band, and \citet{Galaxy_LF_in_UV_at_Cosmic_High_Noon-2017-ApJ}, \citet{Galaxy_primeval_LF_in_UV-2017-arXiv}, etc. use the UV bands), but the central theme is similar for all of the objects -- to measure the intrinsic properties of the sources and statistically study their cosmological evolution. Moreover, the LF of the various objects are related to each other, making this a difficult quantity to measure. For example, the cosmic star-formation rate (CSFR) derived from the galaxy LF, and the GRB LF, are related via Equations \ref{eq:definition_of_phi} and \ref{eq:R_dot}. This will be discussed in more details below.

The measurement of the redshift (hence distance) of a GRB is essential for measuring its intrinsic luminosity. In the era of the Burst and Transient Source Experiment (BATSE) on board the \emph{Compton Gamma Ray Observatory} (CGRO), which detected around $2700$ GRBs in a span of $9$ years (approximately one GRB per day, see \href{https://heasarc.gsfc.nasa.gov/docs/cgro/batse/}{https://heasarc.gsfc.nasa.gov/docs/cgro/batse/}), the measurement of redshift of a detected GRB depended on coincident detection by other instruments with greater localization capabilities. In 1997, the Italian-Dutch satellite BeppoSAX provided the redshift of a burst for the first time via afterglow observations, that of GRB970508 (see \citet{Costa-1997-Natur}, \citet{Bloom_et_al.-1998-ApJ},
\citet{Fruchter_et_al.-2000-ApJ} and references therein). However, the number of GRBs with redshifts measured by BeppoSAX remained only a handful over the years \citep{Amati_et_al.-2002-A&A}. The situation changed entirely with the advent of the Burst Alert Telescope (BAT) on board \s\, \citep{Barthelmy_et_al._2005}, launched in 2004. In addition to detecting roughly $1$ GRB every $3$ days, it has fast on board algorithms to localize the burst and follow it up swiftly with other on-board instruments, the X-Ray Telescope (XRT) and UltraViolet/Optical Telescope (UVOT), as well as other ground-based missions. This provides redshifts via emission lines, absorption lines and photometry of the host-galaxies and/or afterglow, for roughly $\frac{1}{3}^{{\rm rd}}$ of the \s\, GRBs, making it possible to study the intrinsic properties of $\sim300$ GRBs till date (\href{https://swift.gsfc.nasa.gov/archive/grb_table/}{https://swift.gsfc.nasa.gov/archive/grb\_{}table/}).

\citet{Yonetoku_et_al.-2004-ApJ} used the measured redshift and spectral parameters of 12 \emph{BeppoSax} GRBs from \citet{Amati_et_al.-2002-A&A} and an additional 11 GRBs detected by BATSE to derive the `Yonetoku correlation' between the 1-sec peak luminosity and the spectral energy break in the source frame. Using this correlation, they estimated the `pseudo redshift' of 689 BATSE long GRBs with unknown redshifts. Subsequently, they discussed the GRB formation rate and found that a constant LF does not fit the data.

\citet{Daigne_et_al.-2006-MNRAS} studied the logN-logP diagram of BATSE GRBs and the peak-energy distribution of bright BATSE and HETE-2 GRBs, as well as carried out extensive simulations for \s\, GRBs and applied them to early \s\, data to predict that long GRBs show significant cosmological evolution. \citet{Salvaterra_et_al.-2007-ApJ} and \citet{Salvaterra_et_al.-2009-MNRAS} investigated the peak-flux distribution of BATSE GRBs in different scenarios regarding the CSFR, the evolution of the GRB LF, and the metallicity of the GRB formation environments. They then compared the predicted peak-flux distribution of \s\, GRBs primarily with $z>2$ with available data to conclude that the two satellites observe the same distribution of GRBs, the GRB LF shows significant cosmological evolution, and that the GRB formation is limited to low metallicity environments.

Since then, \s\, GRBs with measured redshifts have been studied extensively to model the long GRB LF. To do this, \citet{Wanderman_&_Piran-2010-MNRAS} directly inverted the observed luminosity-redshift relationship. \citet{Cao_et_al.-2011-MNRAS} carried out a phenomenological study of the observational biases on doing this, concluding that a broken powerlaw model of the long GRB LF is preferred, with pre and post break luminosity of $2.5\times10^{52}\,{\rm erg.s^{-1}}$ given by $1.72$ and $1.98$ respectively. They also point to the requirement of cosmological evolution of the LF at high metallicity environments. \citet{Salvaterra_et_al.-2012-ApJ} used a flux-complete sample of 58 \s\, GRBs, with a redshift completeness of $90\%,$ to conclude that the broken powerlaw model is degenerate with the exponential cut-off powerlaw model. They also conclude that the GRB LF evolves with redshift, claiming that this conclusion is independent of the used model. \citet{Robertson_and_Ellis-2012-ApJ} however used a sample of 112 \s\, GRBs to disfavour strong cosmological evolution of the formation rates of GRBs at $z<4,$ and concluded that the best-fit trend of the evolution strongly over-predicts the CSFR at $z>4$ when compared to UV-selected galaxies, alluding to unclear effects in addition to metallicity and the GRB formation environment. \citet{Howell_et_al.-2014-MNRAS} used two new observation-time relations and accounted for the complex triggering algorithm of \s-BAT to reduce the degeneracy between the CSFR and the GRB LF. They satisfactorily fit a non-evolving LF with a powerlaw broken at $0.80\pm0.40\times10^{52}\,{\rm erg.s^{-1}}$ by pre and post indices of $0.95\pm0.09$ and $2.59\pm0.93$ respectively, while not entirely ruling out the possibility of an evolution in the break luminosity. \citet{Petrosian_et_al.-2015-ApJ} used a sample of 200 redshift measured \s\,\emph{ }GRBs to carry out a non-parametric determination of the quantities related to the CSFR and the GRB LF. They claimed that the LF evolves strongly with $z,$ satisfactorily fit to a broken powerlaw model with pre and post break indices $1.5$ and $3.2$ respectively. They also estimated a GRB formation rate an order of magnitude higher than that expected from CSFR at redshifts $z<1,$ but matching with the CSFR at higher redshifts, contrary to all previous studies. On the other hand, \citet{Deng_et_al.-2016-ApJ} carried out an extensive study of the observational biases on the flux-truncation, trigger probability, redshift measurement, etc. with 258 \s\, GRBs, concluding that it is not possible to argue for a robust cosmological evolution of the long GRB LF. The major limitations in the study of the GRB LF with \s\, GRBs is the narrow energy band of BAT, which does not allow an accurate determination of the spectral parameters of the GRBs, since most of the bursts have spectral cutoffs at energies greater than the BAT high-energy threshold of $150$ keV. The conclusions of several of these studies are moreover in contradiction to each other. Regardless, several authors have discussed the implications of these results in the context of the structure of the GRB jets, for both BATSE (\citet{Firmani_et_al.-2004-ApJ}, \citet{Guetta_Granot_Begelman-2005-ApJ}, \citet{Guetta_Piran_Waxman-2005-ApJ}) and \s\, GRBs \citep{Pescalli_et_al.-2015-MNRAS}. The redshift distribution of \s\, bursts emerging from the study of the LFs, assuming different metallicity environments of GRBs, has been discussed by \citet{Natarajan_et_al.-2005-MNRAS}.

The two major limitations of studies that use GRBs with measured redshifts to constrain the GRB LF are: (1) the number of such available sources is rather small to tightly constrain the LF or statistically answer questions related to its evolution with redshift, leading to a variety of conclusions; (2) the measurement of redshifts always suffers from observational biases. To overcome these limitations, \citet{Lloyd-Ronning_et_al.-2002-ApJ} used 220 BATSE GRBs with redshifts inferred from an empirical luminosity-variability relation \citep{Fenimore_and_Ramirez-Ruiz-2000-arXiv}. This was extended by \citet{Firmani_et_al.-2004-ApJ} who carried out a joint fit of these GRBs along with the observed peak-flux distribution of more than 3300 \emph{Ulysses}/BATSE GRBs. The conclusions always favoured a cosmological evolution of the GRB LF, although the data was not able to distinguish between single powerlaw and double-powerlaw models. \citet{Shahmoradi-2013-ApJ} proposed a multivariate log-normal distribution which he fitted for 2130 BATSE GRBs. \citet{Kocevski_&_Liang-2006-ApJ} on the other hand used an empirical lag-luminosity correlation to constrain the GRB LF and the CSFR independently from the study of 900 GRBs, favouring a single powerlaw fit to the GRB LF. Incidentally, similar methods have also been applied for galaxies to study the galaxy LF (see \citet{Galaxy_LF_pseudo_redshift-2017-arXiv} and references therein).

\citet{Tan_et_al.-2013-ApJL} uses the Yonetoku correlation to estimate the pseudo redshifts of 498 GRBs. This avoids the observational bias of the redshift measurements, and the flux truncation is corrected for during the modelling. First they test the correlation parameters by comparing the redshift distribution of 172 \s\, GRB whose redshifts are known. They find that the best-fit parameters do not predict the redshift distribution of this sub-sample well. So they choose the values for which the distributions of known and pseudo redshifts of these GRBs are statistically similar. Since the \s\, bandpass is too narrow to determine the spectral parameters of \s\, GRBs, they use the \citet{Butler_et_al.-2007-ApJ} catalog in which the Band function \citep{Band_et_al.-1993-ApJ} parameters are estimated with a Bayesian technique. They conclude that the GRB LF is inconsistent with a simple powerlaw, demanding a fit with a broken powerlaw with pre and post break indices given by $0.8$ and $2.0$ respectively. In addition, the break itself evolves cosmologically as $L_{b}=1.2\times10^{51}\,{\rm erg.s^{-1}}(1+z)^{2},$ and the GRB formation rate evolves as $\propto(1+z)^{-1}$ in contradiction to all previous studies. They do not look into the accuracy of pseudo redshifts of the GRBs individually, and the analysis is entirely based only on a statistical comparison of the redshift distributions.

In the present work, I carry out a detailed study of the estimation of pseudo redshifts, using long GRBs that have firm redshift measures from \s, as well as spectral parameter measurements from \f. The reason such a sample is useful is because it combines the wide spectral coverage of \f\, (which however does not provide redshift) with the redshift measurements from \s\,\emph{ }follow-ups. This reduces the errors on the correlated quantities compared to the Butler catalog, which allows me to test the correlation itself, and also carefully examine the accuracy of the pseudo redshifts estimated from the correlation. I then use it to estimate the pseudo redshift  of all \f\, and \s\, GRBs, and place constraints on the long LF from a combined study of all these $2067$ GRBs. Previously, \citet{Yu_et_al.-2015-ApJS} has used a combined sample of 127 long GRBs with spectra from \f\, and Konus-\emph{Wind}, and redshift from \s, to independently model the CSFR and the GRB LF. They used the GRBs irrespective of whether the spectral peak is actually seen in the instrumental waveband. In the present work, I choose only those bursts in which the spectral peak is accurately modelled, to re-derive the parameters of the Yonetoku correlation, which is then used to include a much larger number of sources.

This paper is organized as follows. In Section \ref{sec:The-Yonetoku-correlation}, the Yonetoku correlation is re-derived. In Section \ref{sec:The-estimated-luminosities}, I describe the use of the correlation to generate pseudo redshifts of all remaining \f\, and \s\, GRBs. The GRB LF is modelled in Section \ref{sec:Modeling-the-GRB-LF}, and in Section \ref{sec:Conclusions}, I present concluding remarks. Throughout this paper, a standard $\Lambda$CDM cosmology with $ H_0 = 72 \, \rm{km.s^{-1}.Mpc^{-1}} ,$ $ \Omega_m = 0.27 $ and $ \Omega_{\Lambda} = 0.73 $ has been assumed. All the scripts used and important databases generated in the work are publicly available at \url{https://github.com/DebduttaPaul/luminosity_function_of_LGRBs_using_Swift_and_Fermi}.

\section{The Yonetoku correlation}
\label{sec:The-Yonetoku-correlation}

It is the correlation seen between the peak luminosity $L_{p}$ and the spectral energy break $E_{p}$ \citep{Band_et_al.-1993-ApJ} in the source frame.

The peak luminosity is defined as

\begin{equation}
L_{p}=P.\,4\pi d_{L}(z)^{2}\times k(z;\,{\rm spectrum}),\label{eq:Luminosity_formula}
\end{equation}

\noindent where $P$ denotes the peak flux modelled by the Band function during the burst duration, given in ${\rm erg.cm^{-2}s^{-1},}$ and

\begin{equation}
k(z)=\dfrac{\int_{1\,{\rm keV}}^{10^{4}\,{\rm keV}}E.S(E)dE}{\int_{(1+z)E_{min}}^{(1+z)E_{max}}E.S(E)dE}
\label{eq:definition_of_k--Fermi}
\end{equation}
for \f\, GRBs. In case of \s\, bursts, where the peak flux is given in ${\rm ph.cm^{-2}s^{-1},}$ 

\begin{equation}
k(z)=\dfrac{\int_{1\,{\rm keV}}^{10^{4}\,{\rm keV}}E.S(E)dE}{\int_{(1+z)E_{min}}^{(1+z)E_{max}}S(E)dE}\,.\label{eq:definition_of_k---Swift}
\end{equation}

\begin{figure}
\centering{}\includegraphics[scale=0.28]{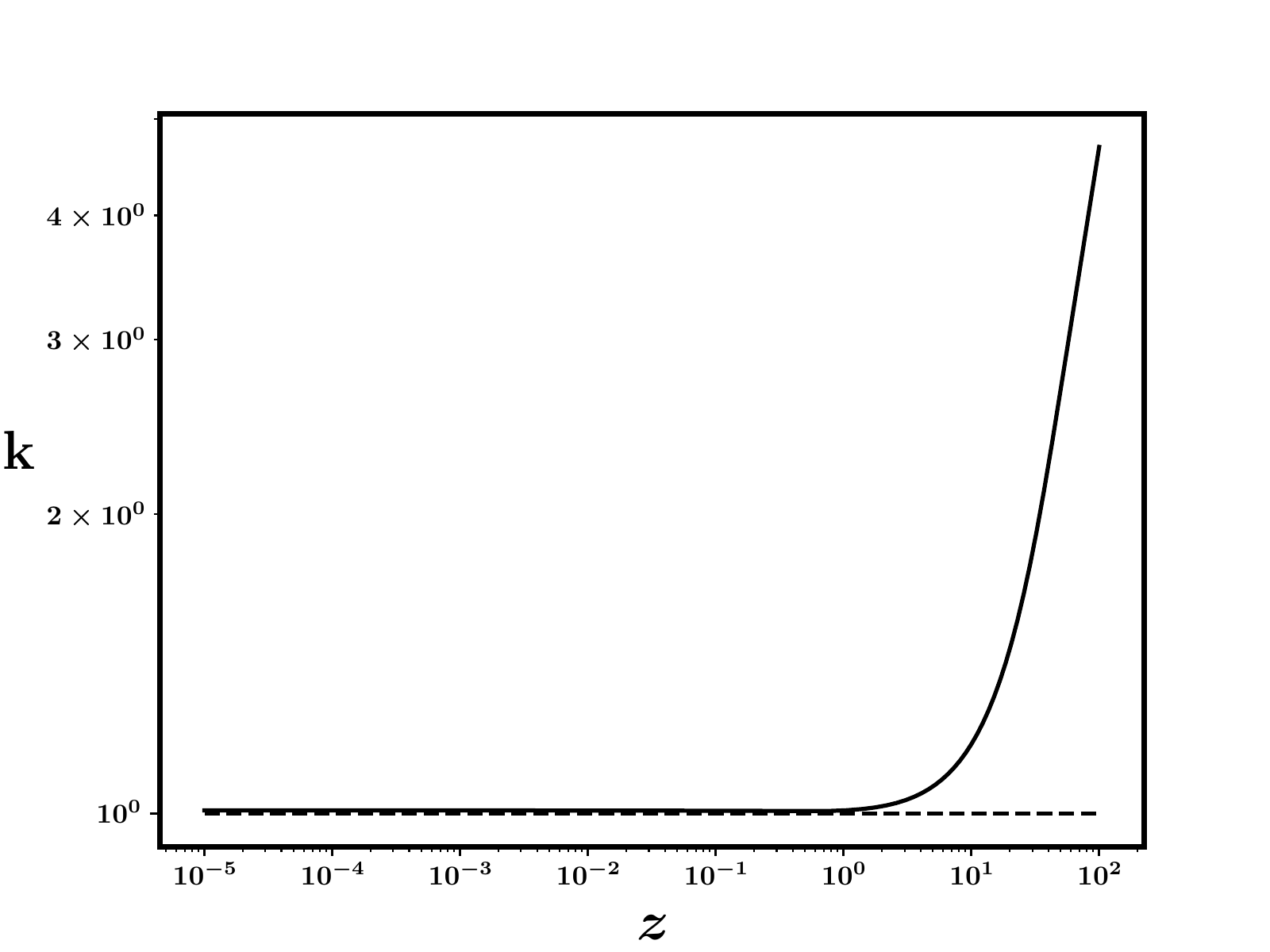}\includegraphics[scale=0.28]{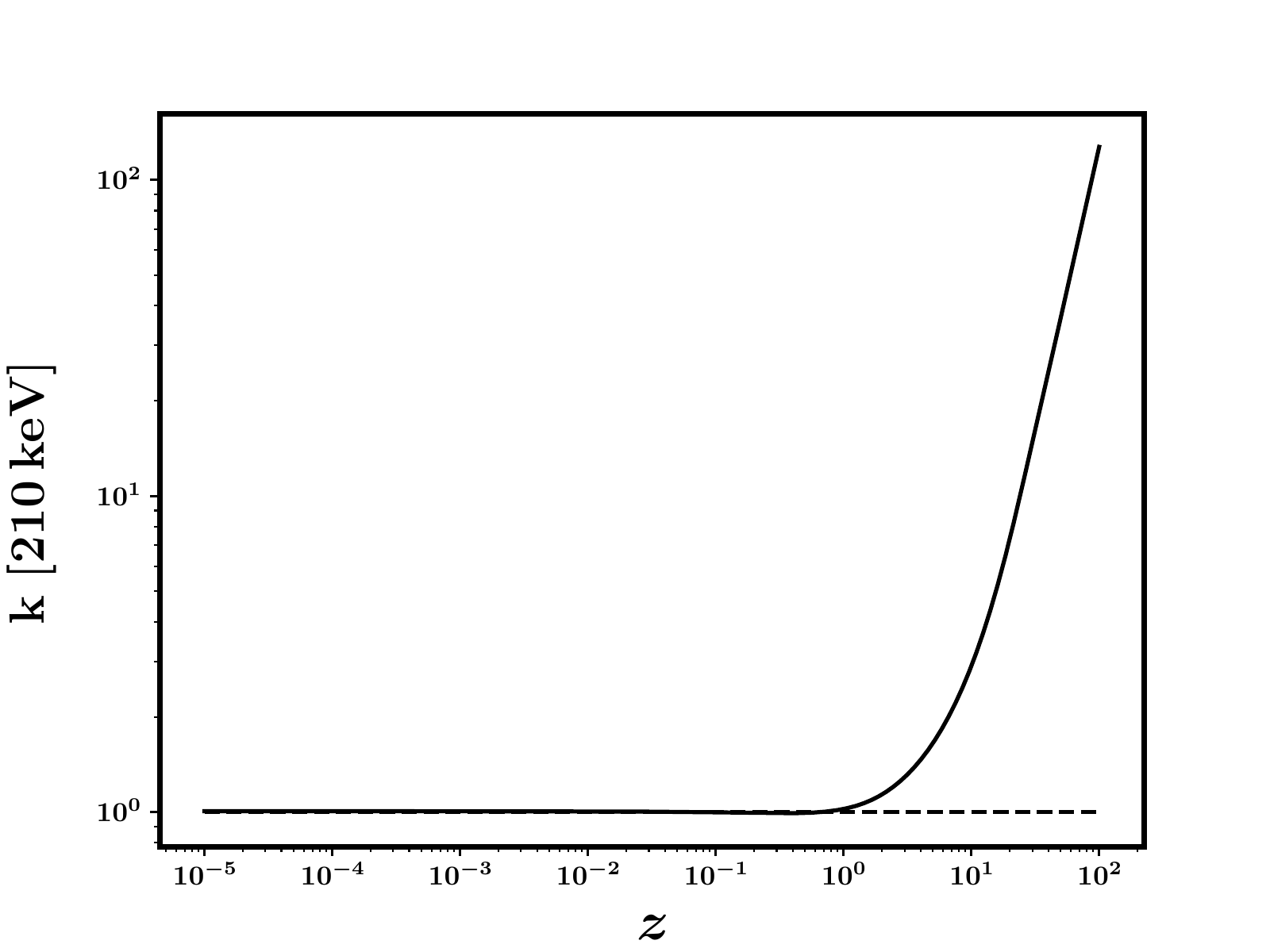}\caption{The $k$-correction factors for \f\, (left) and \s\, (right) assuming average spectral parameters as derived from the sample of \f\, bursts: $<E_{p}>\,=181.3$ keV,
$<\alpha>\,=-0.566,$ $<\beta>\,=-2.823.$ Since these average numbers are used, they do not include uncertainties. Note that the units of $k$ are different for \f\, and \s, owing to Equations \ref{eq:definition_of_k--Fermi} and \ref{eq:definition_of_k---Swift}. One also notices the striking difference in the scales: whereas it is much close to unity for \f\, which is a wide-band detector, for \s\, it is much larger for large redshifts than its value at the local universe, because of \s's limited energy-range. \label{fig:k-correction}}
\end{figure}

To accurately derive the Yonetoku correlation, I first select the sub-sample of all \f\, and \s\, bursts that have accurate estimations of the Band function \citep{Band_et_al.-1993-ApJ} parameters during the prompt emission, by \f, as well as accurate redshift measurement by \s\, follow-up. Previous works have relied on modeling the spectral parameters by \s, which suffers from the limited wavelength range of BAT. I use the accurate spectral parameters from \f\, instead, reducing the inaccuracy of the estimates of luminosity. Moreover, due to the same reason, I also notice that the $k$ correction is very close to unity for these bursts, unless the redshift is not too large (even for $z=10,$ the factor is less than $1.5$). This is illustrated in left of Fig \ref{fig:k-correction}. In comparison, the k-correction of \s\, is much larger for larger redshifts.

\subsection{Selecting the common GRBs}

\begin{figure}
\centering{}\includegraphics[scale=0.5]{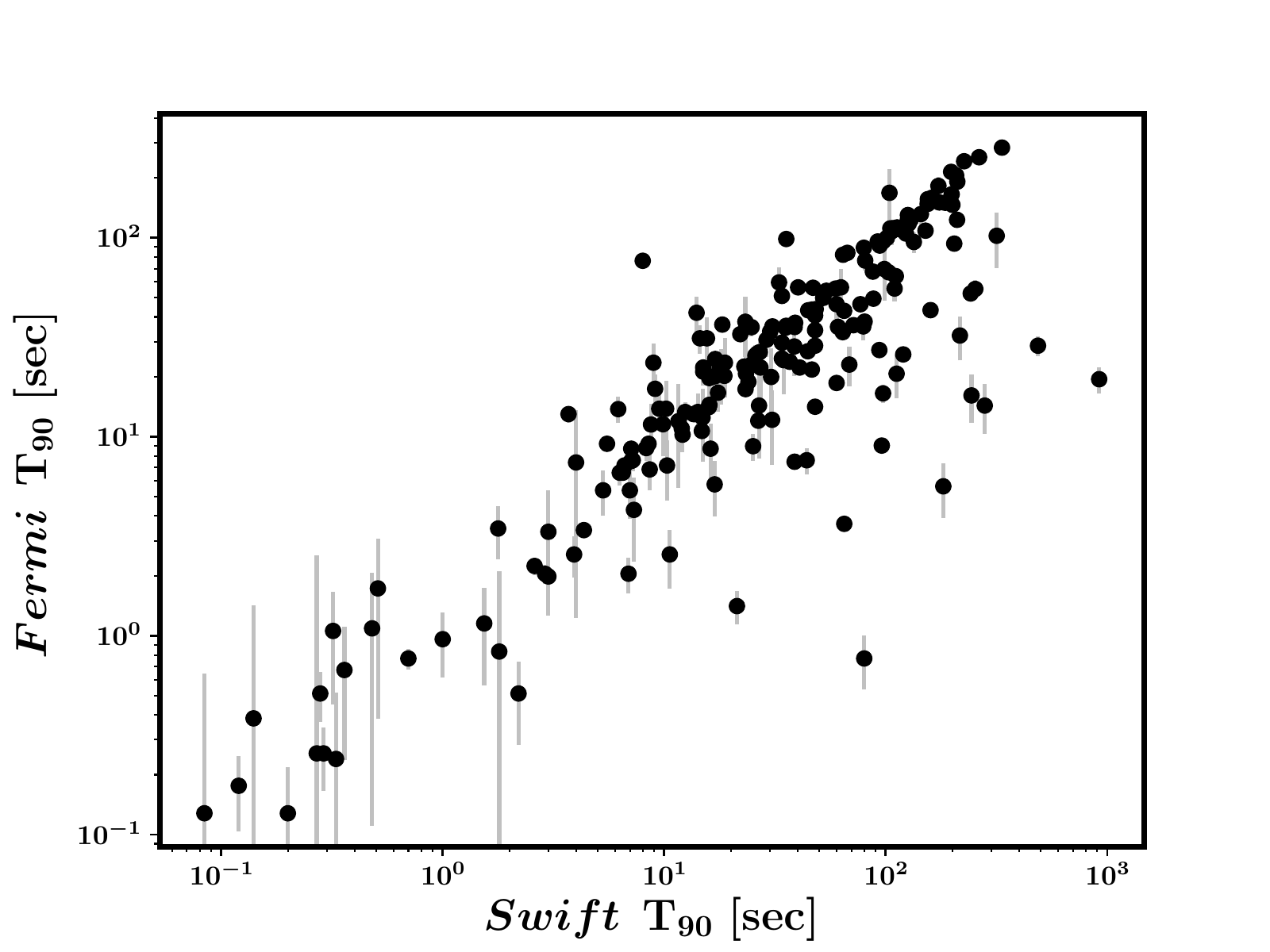}\caption{Despite the expected correlation between the \f\, and \s\, $\T$s, some GRBs have systematically smaller $\T$ in \f\, than \s.\label{fig:T90_comparison}}
\end{figure}

The updated list of \f\, GRBs are selected from the \f\, catalog\footnote{\href{https://heasarc.gsfc.nasa.gov/W3Browse/fermi/fermigbrst.html}{https://heasarc.gsfc.nasa.gov/W3Browse/fermi/fermigbrst.html}} till GRB170501467, which includes 2070 GRBs. Firstly, I choose only
those bursts from the catalog that have spectroscopically measured
parameters for the GRB Band function, which includes 1729 such cases.
Then only those with small errors on the spectral parameters are chosen. For this, it is noted that the primary parameter that drive the error estimates in the luminosity is the $\Ep.$ Choosing only those with errors less than $100\%$ in $E_p,$ 1566 bursts are retrieved.

The updated list of \s\, GRBs are selected from the \s\, catalog\footnote{\href{https://swift.gsfc.nasa.gov/archive/grb_table/}{https://swift.gsfc.nasa.gov/archive/grb\_{}table/}} till GRB 170428A. The total number is 1021, out of which those with
firm redshift measure are 312.

Since the nomenclature of \f\, and \s\, GRBs are different, I select the following criteria for selecting the common GRBs. The difference between the trigger times are selected to be less than 10 minutes, and they are restricted to within $10^{\circ}\times10^{\circ}$ in RA and Dec for the two instruments. These numbers are empirically chosen, such that the common number of GRBs converge within a reasonable range of these cutoffs. This ensures I do not mistake two GRBs which are well separated in time and space to be the same GRB. Consequently I get 68 common GRBs. Applying the $\T$ criterion for identifying short versus long bursts \citep{Kouveliotou_et_al.-1993-ApJ} separately for the two missions, I note that 65 are long according to both \f\, and \s, two are short in both, while only one is short only in \f, GRB090927422 (\f\, nomenclature). Its \f-$\T$ is $0.512\pm0.231$ sec while that of \s\, is $2.2$ sec. \f-$\T$s are calculated at higher energies and hence known to be systematically smaller in a handful of GRBs. Fig \ref{fig:T90_comparison} illustrates this effect. Hence, I choose this as a long burst. Moreover, this also gives me confidence to make the distinction between long and short GRBs based on the \s-criterion whenever it is available, i.e. for the other common GRBs (without redshift estimates from \s). For the ones that are detected only by \f, I resort to applying the criterion based on the \f-$\T.$

\subsection{Testing the correlation}

\begin{figure}
\centering{}\includegraphics[scale=0.5]{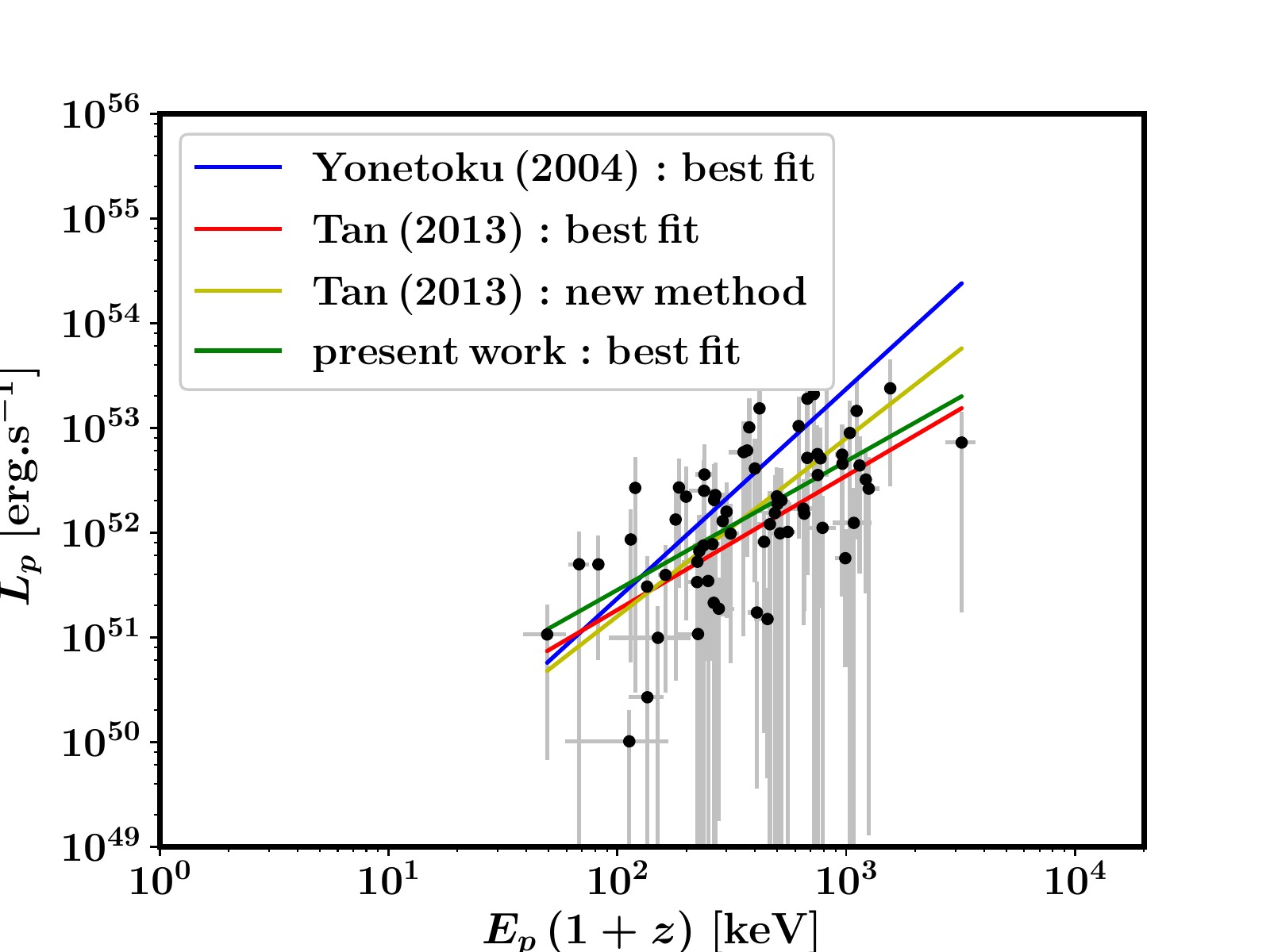}\caption{The Yonetoku correlation as seen from the data of 66 long GRBs with accurate Band parameters from \f\, and redshift measurement from \s. The parameters of the correlation, from various studies, are over-plotted. I get the best-fit parameters of $A=4.783\pm1.026$ and $\eta=1.227\pm0.038$ for the the correlation defined in Equation \ref{eq:Yonetoku_correlation}. \newline{}(A coloured version of this figure is available in the online journal.)
\label{fig:Yonetoku_correlation}}
\end{figure}

\begin{figure}
\centering{}\includegraphics[scale=0.28]{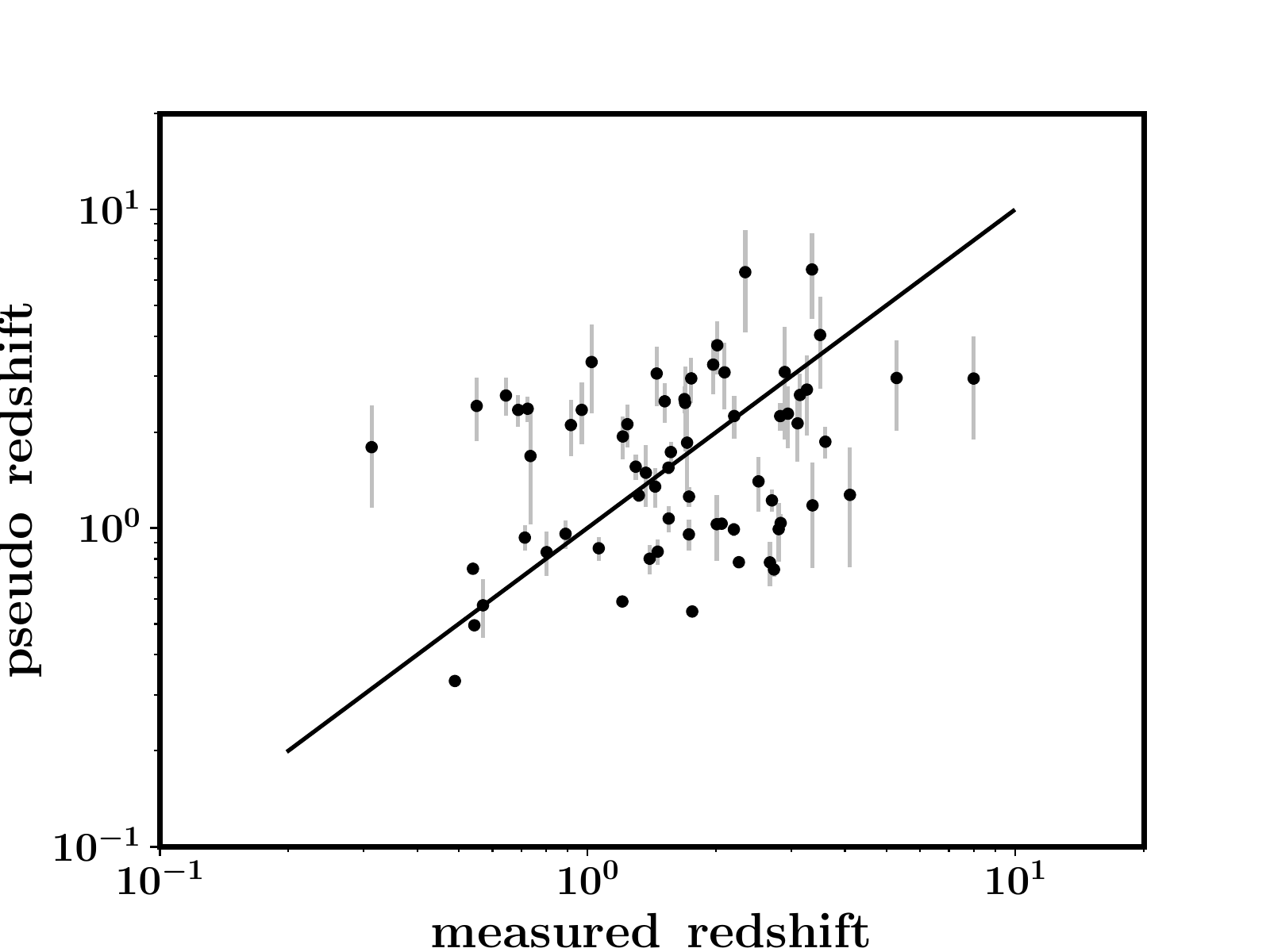}\includegraphics[scale=0.28]{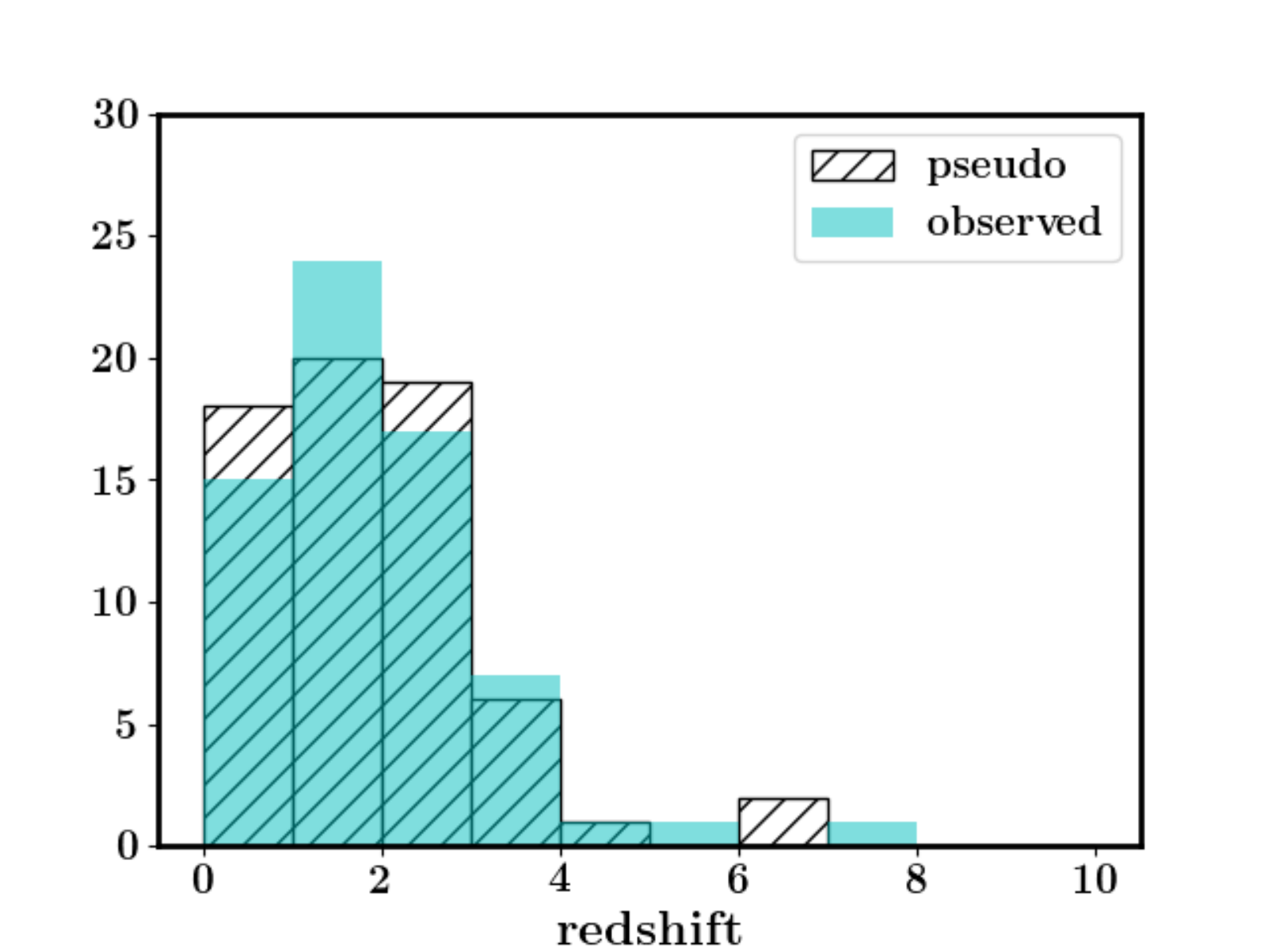}\caption{The redshift distribution for the 66 long GRBs chosen in our sample. \emph{Left:} Individual comparison, the line indicating the expected relationship if the method was successful in predicting the pseudo redshifts accurately. \emph{Right:} Statistical comparison: filled (cyan) histogram shows the observed distribution, hatched (black) histogram shows the pseudo redshift distribution. The small discrepancies, specially at higher redshifts, can be easily understood to be due to the errors on the pseudo redshifts. \newline{}(A coloured version of this figure is available in the online journal.) \label{fig:redshift_distribution--bestfit}}
\end{figure}

When I plot $L_{p}$ versus $E_{p}(1+z)$ (the factor of $(1+z)$ takes care of the transformation into the co-moving frame) for all the 68 GRBs, I notice that the only burst with systematically smaller $L_{p}$ than the rest, is a short burst. Moreover, the sample of short GRBs with accurate spectral and redshift measures consists of only two cases. Hence, I do not attempt to study the correlation for short bursts separately. Moreover, I do not find any burst with luminosity lower than $ 10^{49} \rm{ \, erg.s^{-1} }, $ nor with $\T > 10^3 \, \rm{sec}, $ and hence I do not attempt to segregate the possible separate classes of low-luminosity long GRBs (see e.g. \citet{Liang_et_al.-2007-ApJ}), or ultra-long GRBs (e.g. \citet{Levan_et_al.-2014-ApJ}).

I retrieve the Yonetoku correlation from the $66$ long bursts to a high degree of confidence (a null-hypothesis of the Spearman correlation co-efficient of $0.623$ being false, ruled out with $p=2.368\times10^{-8}$), as shown in Fig \ref{fig:Yonetoku_correlation}. The errors on $L_{p}$ consist of errors in the flux as well as a conservative estimate of $70\%$ systematic error added to all bursts, to take care of the inaccuracy in the spectral parameters. These parameters are non-linear and hence the errors cannot be calculated directly. The systematic error is chosen conservatively, since the changes in the spectral parameters always affect the estimates in $L_{p}$ within a factor of $1.5$ even for the highest redshift bursts (see Fig \ref{fig:k-correction} for reference). Also, if linear errors are propagated, the mean errors are again of the same order.

\begin{figure}
\centering{}\includegraphics[scale=0.5]{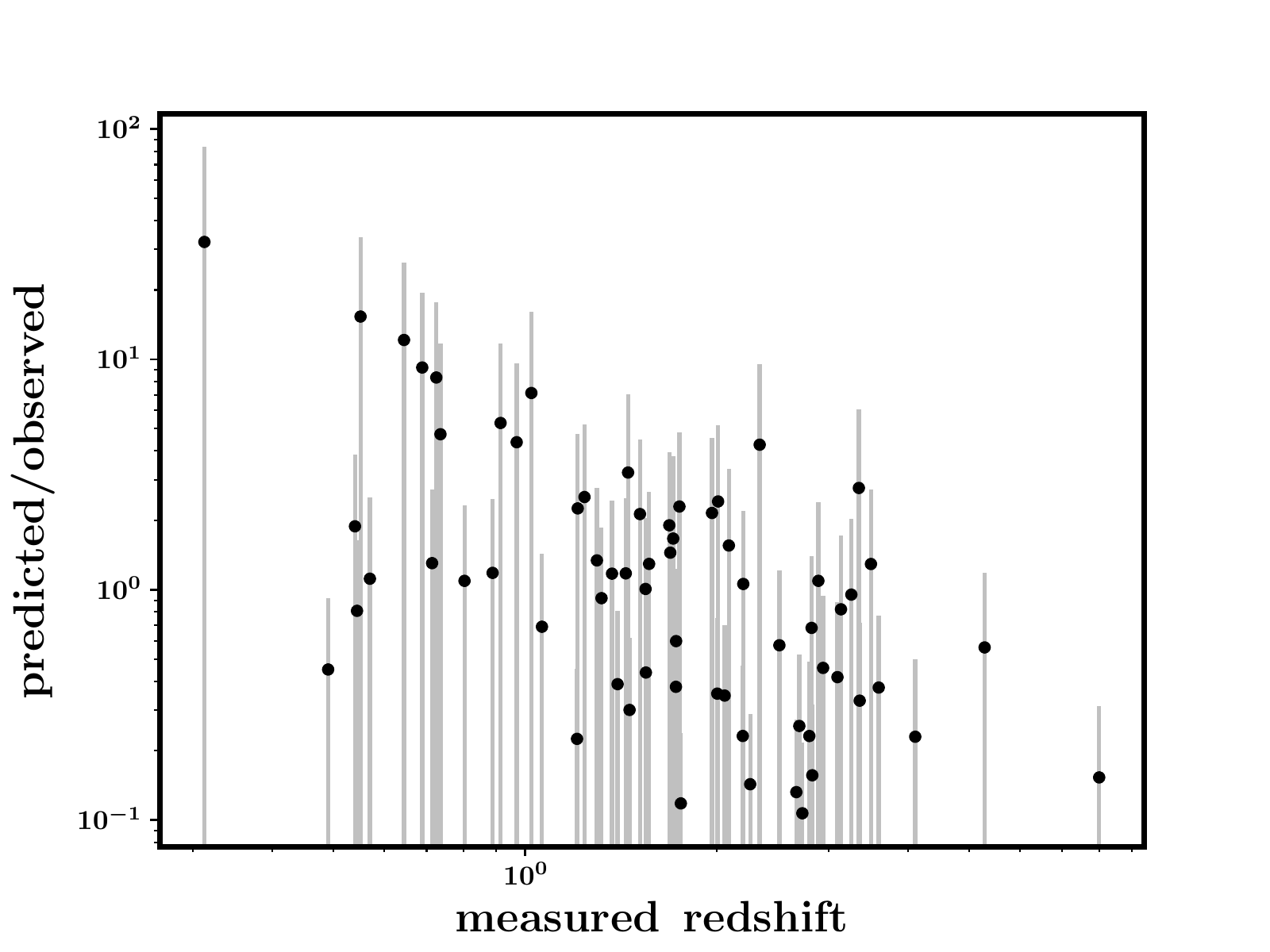}\caption{A strong anti-correlation is seen against the measured redshift, for the ratio between the luminosities predicted (from the best-fit Yonetoku correlation) and that measured directly.\label{fig:correlation_of_ratio_with_measured_z}}
\end{figure}

For the Yonetoku correlation defined as 
\begin{equation}
\dfrac{L_{p}}{10^{52}{\rm \, erg.s^{-1}}}=A.\left[\dfrac{E_{p}}{{\rm MeV}}(1+z)\right]^{\eta},\label{eq:Yonetoku_correlation}
\end{equation}
I get the best-fit parameters of $A=4.780\pm0.123$ and $\eta=1.229\pm0.037.$ The corresponding redshift distributions for the same GRBs, both statistically and individually, are shown in Fig \ref{fig:redshift_distribution--bestfit}. It is noticed that although the method does not reproduce the redshifts on an individual basis, it is statistically reliable. The pseudo and observed redshifts has a median ratio of $1.002 \pm 0.721$, i.e. the number is consistent with unity. This is not an effect of normalization, as all the normalization factors are defined explicitly via Equation \ref{eq:Yonetoku_correlation}. The reason of it being statistically reliable is that, the method produces the pseudo redshifts of a larger sample by assuming gross parameters from a smaller sample which is however unbiased. The systematic discrepancies for individual bursts can be ascribed to the scatter around the Yonetoku correlation, as discussed below.

\citet{Tan_et_al.-2013-ApJL} uses the set of parameters that reduce the discrepancy between the distributions of the observed and pseudo redshifts. This method tries to reconcile the problem by changing the parameters, while circumventing the actual problem, that the Yonetoku correlation is intrinsic scattered. This is best illustrated by the left panel of Fig \ref{fig:redshift_distribution--bestfit}. Moreover to verify their method, I run it on the current dataset, to find no global minimum of the discrepancy between the distributions. Hence, instead of modifying the parameters, I investigate the possible reasons for the scatter.

To investigate the presence of systematics in the discrepancy between the observed and the pseudo redshifts, I look for possible correlations of the ratio of the predicted luminosity from the Yonetoku correlation with the physical parameters $E_{p}(1+z)$ and the measured redshift. No correlation is found with the former, which confirms that the scatter in the Yonetoku correlation is intrinsic. However, I find a strong anti-correlation between the ratio and the measured redshift, as shown in Fig \ref{fig:correlation_of_ratio_with_measured_z}, with a null hypothesis of the Spearman correlation co-efficient of $-0.533$ being false, ruled out with $p=4.056\times10^{-6}.$ The following qualitative hypothesis is proposed to explain this trend. The luminosities predicted by the best-fit parameters of the observed correlation are the better physical estimates of the luminosity, physically correlating with the spectral peak. The scatter in the observed correlation between the quantities $L_{p}$ and $E_{p}$ (in the source frame) is due to the inadequacy of the definition of the luminosity, which needs to be corrected for physical factors like the beaming of the burst and the burst environment. This explanation, however, is qualitative and requires an in-depth analysis via modeling the possible physical effects, not attempted in the current work.

\section{The estimated luminosities}
\label{sec:The-estimated-luminosities}

I next calculate the luminosities of all the \f\, detected bursts. This includes the 66 GRBs already used in Section \ref{sec:The-Yonetoku-correlation}, and the rest with spectral estimates from \f\, but without redshift estimates from \s\, (irrespective of they are detected by \s). For the latter cases, pseudo redshifts are predicted via the Yonetoku correlation, using \f\, flux and $k$-corrections. However the \s-$\T$ criterion is applied to those with \s-detections to distinguish between the short and long classes. For the GRBs with only \s\, detections along with measured redshifts, I directly calculate the luminosity from the flux and redshifts from the same catalog, and the \s\, $k$-corrections derived from the Band function parameters fixed at the average values of the \f\, distribution, given by $<E_{p}>\,=181.3$ keV, $<\alpha>\,=-0.566,$ $<\beta>\,=-2.823.$ It is to be noted that the $k$-correction is not sensitive to these parameters, as long as they are within a reasonable range (see e.g. \citet{Preece_et_al.-2000-ApJS} for the study of BATSE bursts). For those bursts detected only by \s\, and further lacking redshift measurements, I estimate the pseudo redshifts via the \s\, $k$-corrections and the Yonetoku correlation. Since $E_{p}$ features explicitly in the correlation, they are randomly sampled from the distribution of the \f\, bursts. The justification for such an approach is again that the \f\, being a wide-band detector, samples out all possible values of $E_{p}.$

In Fig \ref{fig:pseudo_redshifts_and_luminosities} is shown the $L$-$z$ distribution of all these cases. The instrumental sensitivities are given by Equation \ref{eq:Luminosity_formula} with $P=8.0\times10^{-8}\,{\rm erg.cm^{-2}.s^{-1}}$ for \f\, and $P=0.2\,{\rm ph.cm^{-2}.s^{-1}}$ for \s\, (for a $100$ keV photon, this is equivalent to $3.2\times10^{-8}\,{\rm erg.cm^{-2}.s^{-1}}$). These numbers are chosen empirically from the respective catalogs, and describe the lower cutoff well. This places confidence on the used method and the estimated luminosities, and I proceed to use them for modeling the luminosity function (in Section \ref{sec:Modeling-the-GRB-LF}). The slopes of the two correlations are $1.584\pm0.002$ for \f\,and $1.834\pm0.002$ for \s. A few bursts (eight) fall below the sensitivity line, which may be ascribed to the fact that the spectral parameters are sampled randomly from the \f\, distribution, whereas the flux is measured by \s; also, the $k$-correction increases sharply with $z$ for \s. These bursts are removed from the sample for subsequent analysis.

\begin{table}

\caption{The type of \f\, and \s\, long GRBs used for modeling, and how they are referred. The total number is 2067.\label{tab:GRB_numbers}}

\begin{centering}
\begin{tabular}{|c|c|c|c|}
\hline 
type & redshift measured & number & modelled as\tabularnewline
\hline 
\hline 
both \f\, and \s & yes & 66 & \multirow{2}{*}{\f}\tabularnewline
\cline{1-3} 
only \f, or both & no & 1278 & \tabularnewline
\hline 
only \s & no & 499 & \multirow{2}{*}{\s}\tabularnewline
\cline{1-3} 
only \s & yes & 224 & \tabularnewline
\hline 
\end{tabular}
\par\end{centering}

\end{table}

On an average, the pseudo redshifts have $ \sim 20 \% $ errors and the luminosities calculated from them have $ \sim 40 \% $ errors, after propagating errors in all the estimation steps. Theoretically, the redshifts and hence luminosities of the \s\, bursts have much larger uncertainties, because their $E_p$s are not known, but this fact is ignored, to use these bursts in the statistical sense, laying no claim to the accuracy of the individual pseudo redshifts.

I also note that the distribution of pseudo redshifts and corresponding luminosities are relatively insensitive to the exact value of the parameters used for the Yonetoku correlation, as long as they are not significantly different from the best-fit estimates. The advantage of using this method lies in the fact that it evades the complex observational biases that plague and limit the study of redshift measured bursts. Also, it allows the model to take care of the instrumental thresholds while modeling the luminosity function via Equation \ref{eq:definition_of_phi}, to which I turn next.

\section{Modeling the long GRB luminosity function}
\label{sec:Modeling-the-GRB-LF}

\begin{figure}
\centering{}\includegraphics[scale=0.28]{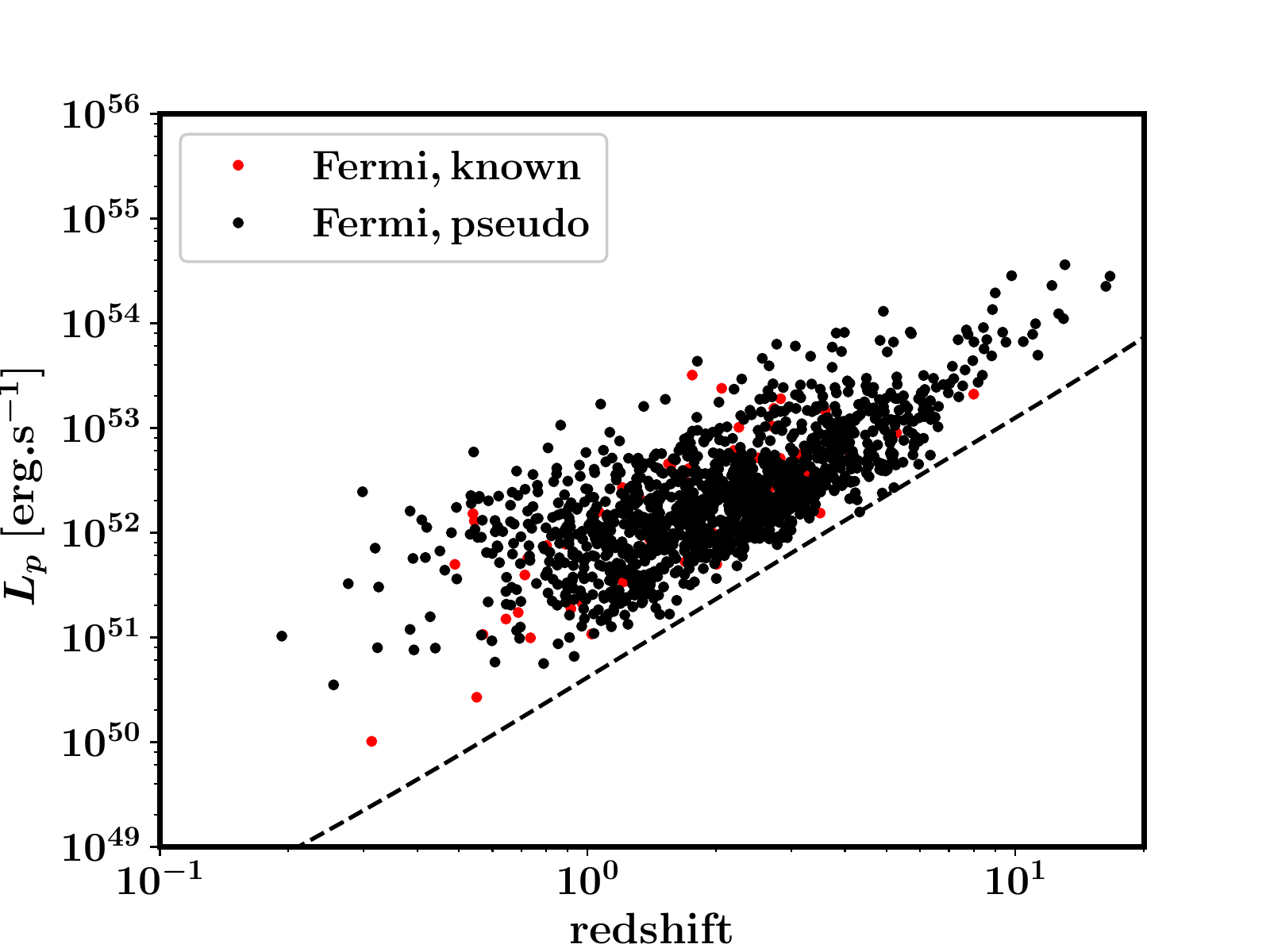}\includegraphics[scale=0.28]{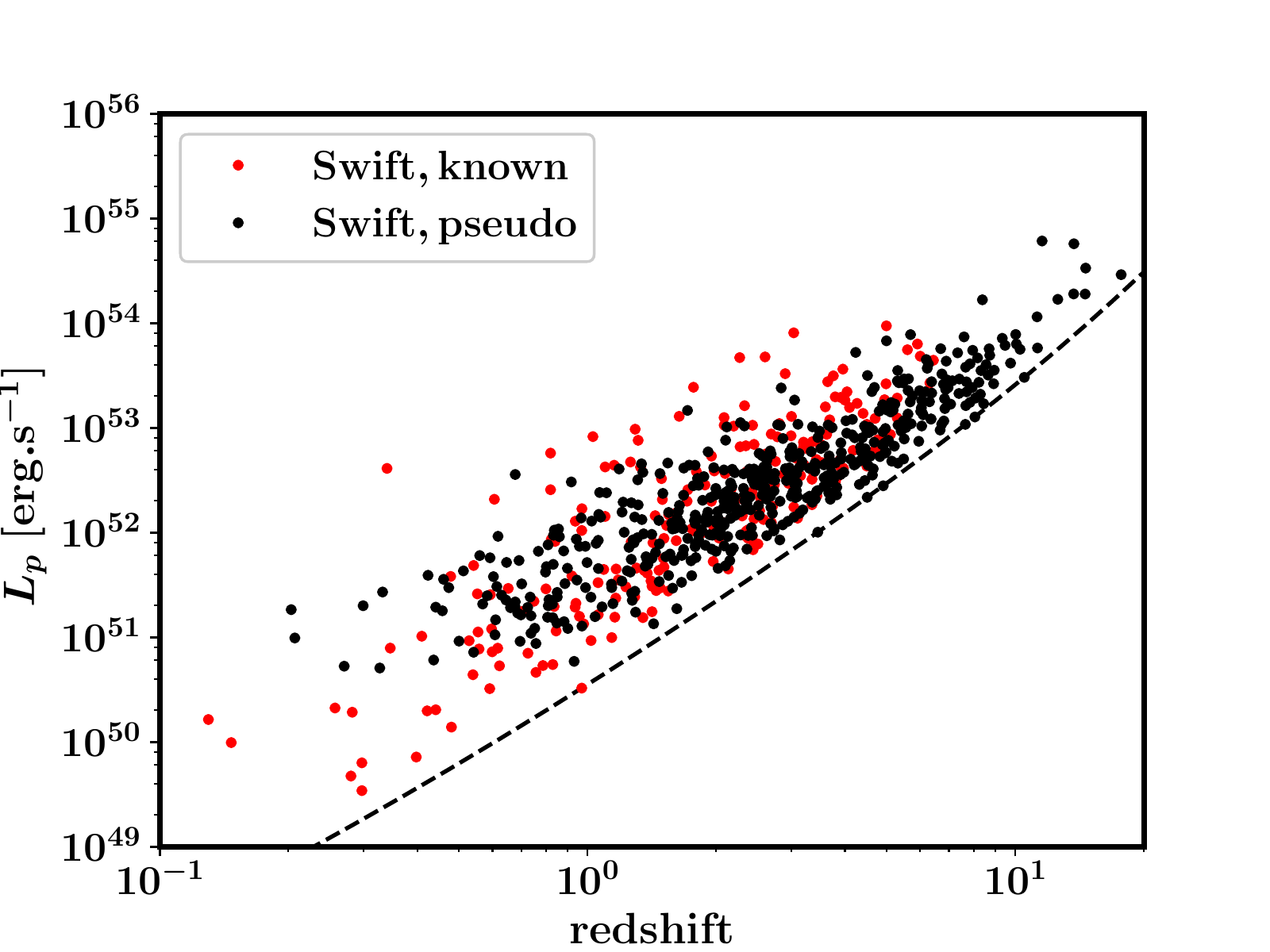}\caption{The luminosity versus redshifts of all GRBs. The red points are for those with redshift measurements, while black points are for those whose pseudo redshifts are derived as described in the text. The dotted lines show the corresponding instrumental sensitivity limits. The errors are not shown for the purpose of better visibility. \emph{Left:} For the GRBs that are detected by \f, irrespective of \s\, detection, including those with known redshifts (the 66 cases considered to study the correlation). For all these bursts, the \f\, $k$-correction is used, whereas the \s-$\T$ criterion is applied for those available. \emph{Right:} For the bursts with detection only by \s, including those with measured redshifts. See Table \ref{tab:GRB_numbers} for more details on the nomenclature. \newline{}(A coloured version of this figure is available in the online journal.) \label{fig:pseudo_redshifts_and_luminosities}}
\end{figure}

For the purpose of modeling the luminosity function, the GRBs that have pseudo redshift greater than $10$ are not considered (27). The final number of GRBs used are showed in Table \ref{tab:GRB_numbers}. Also, the modeling is carried out separately for \f\, and \s, since the cut-off luminosities which feature in the model, via Equation \ref{eq:definition_of_phi}, are different for the two instruments, as discussed in Section \ref{sec:The-estimated-luminosities}. For each instrument, I bin the data into three equipopulous redshift bins: $0<z<1.538,$ $1.538\leq z<2.657,$ $2.657\leq z<10.0$ for \f, and $0<z<1.809,$ $1.809\leq z<3.455,$ $3.455\leq z<10.0$ for \s. It is to be noted that the errors on $N(L)$ are proportionally large, due to the large percentage errors on the derived luminosities, which are propagated across the bins.

In the most recent work on GRB LF, \citet{Amaral-Rogers_et_al.-2017-MNRAS} discusses various kinds of models. In particular, they test models in which the GRB formation rate is tied to a single population of progenitors via the cosmic star formation rate, another similar but distinct model where low and high luminosity GRBs are separated into two distinct classes, and a third kind where no assumption of the GRB formation rates are made. They conclude that a clear distinction between the three kinds of models cannot be asserted however. In the present work, I do not attempt to classify low and high luminosity GRBs for the reason that there is no clear evidence from the study in Section \ref{sec:The-Yonetoku-correlation}. Moreover, I assume that the GRB formation rate is proportional to the star-formation rate, because after all it is massive stars formed in the galaxies that later end their lives in GRBs. There may be an additional dependence on the redshift: most generally represented via Equation \ref{eq:R_dot}. I take the cosmic star-formation rate $\overset{.}{\rho_{\star}}\left(z\right)$ from \citet{Bouwens_et_al.-2015-ApJ} (see references therein for the values at different redshifts), and model additional dependencies of the normalization, that is the GRB formation rate per unit cosmic star formation rate (or the GRB formation efficiency), as 

\begin{equation}
f_{B}C(z)\propto\left(1+z\right)^{\epsilon}.\label{eq:fB.C_of_z}
\end{equation}

\noindent It is to noted that the detailed processes involved in the formation of GRBs do not affect this treatment, which is similar to that followed by \citet{Tan_et_al.-2013-ApJL}. Within this framework, I attempt to fit two models: the exponential cut-off powerlaw (ECPL) model, described by

\begin{equation}
\Phi_z(L)=\Phi_{0}
\left(\frac{L}{L_{b}}\right)^{-\nu} \exp\left[- \left(\frac{L}{L_{b}}\right) \right],
\label{eq:The-ECPL-model}
\end{equation}

\noindent and the broken powerlaw (BPL) model, given as

\begin{equation}
\Phi_z(L)=\Phi_{0}\begin{cases}
\left(\frac{L}{L_{b}}\right)^{-\nu_{1}}, & L\leq L_{b}\\
\left(\frac{L}{L_{b}}\right)^{-\nu_{2}}, & L>L_{b}.
\end{cases}
\label{eq:The-BPL-model}
\end{equation}

\noindent Moreover, most generally the `break-luminosity' $L_{b}$ is allowed to vary with redshift, as

\begin{equation}
L_{b}=L_{b,0}\left(1+z\right)^{\delta},\label{eq:evolution_of_break_luminosity}
\end{equation}

\noindent with the quantity $L_{b,0}$ describing the normalization at zero redshift, and $\delta$ describing the evolution with redshift. The quantity $\Phi_{0}$ normalizes the probability density function $\Phi(L),$ and is an implicit function of the redshift $z$ via the dependence on $L_{b}.$ The models are then described by Equations \ref{eq:definition_of_phi}, \ref{eq:R_dot}, \ref{eq:fB.C_of_z}, \ref{eq:The-ECPL-model}, \ref{eq:The-BPL-model} and \ref{eq:evolution_of_break_luminosity}, along with $\overset{.}{\rho_{\star}}\left(z\right)$ extracted numerically from \citet{Bouwens_et_al.-2015-ApJ}.

\begin{table}
\caption{The best-fit parameters for the ECPL model, as found by search in the $2$-dimensional space of $\nu$ and $L_{b,0}$. As a comparison, I show the best-fit parameters for the equivalent model of the recent works of \citet{Amaral-Rogers_et_al.-2017-MNRAS}. We see an overall agreement between the values.
\label{tab:ECPL_model_parameters}}
\begin{centering}
\begin{tabular}{|c|c|c|c|}
\hline 
parameter & present work & \citet{Amaral-Rogers_et_al.-2017-MNRAS}\tabularnewline
\hline 
\hline 
$\nu$ & $0.60 \pm 0.1 $ & $0.71 \pm 0.07$\tabularnewline
\hline 
$L_{b,0}$ & $5.40_{-1.5}^{+2.0}$ & $4.02_{-0.96}^{+1.52}$\tabularnewline
\hline 
\end{tabular}
\par\end{centering}
\end{table}

\begin{figure*}
\begin{centering}
\includegraphics[scale=0.39]{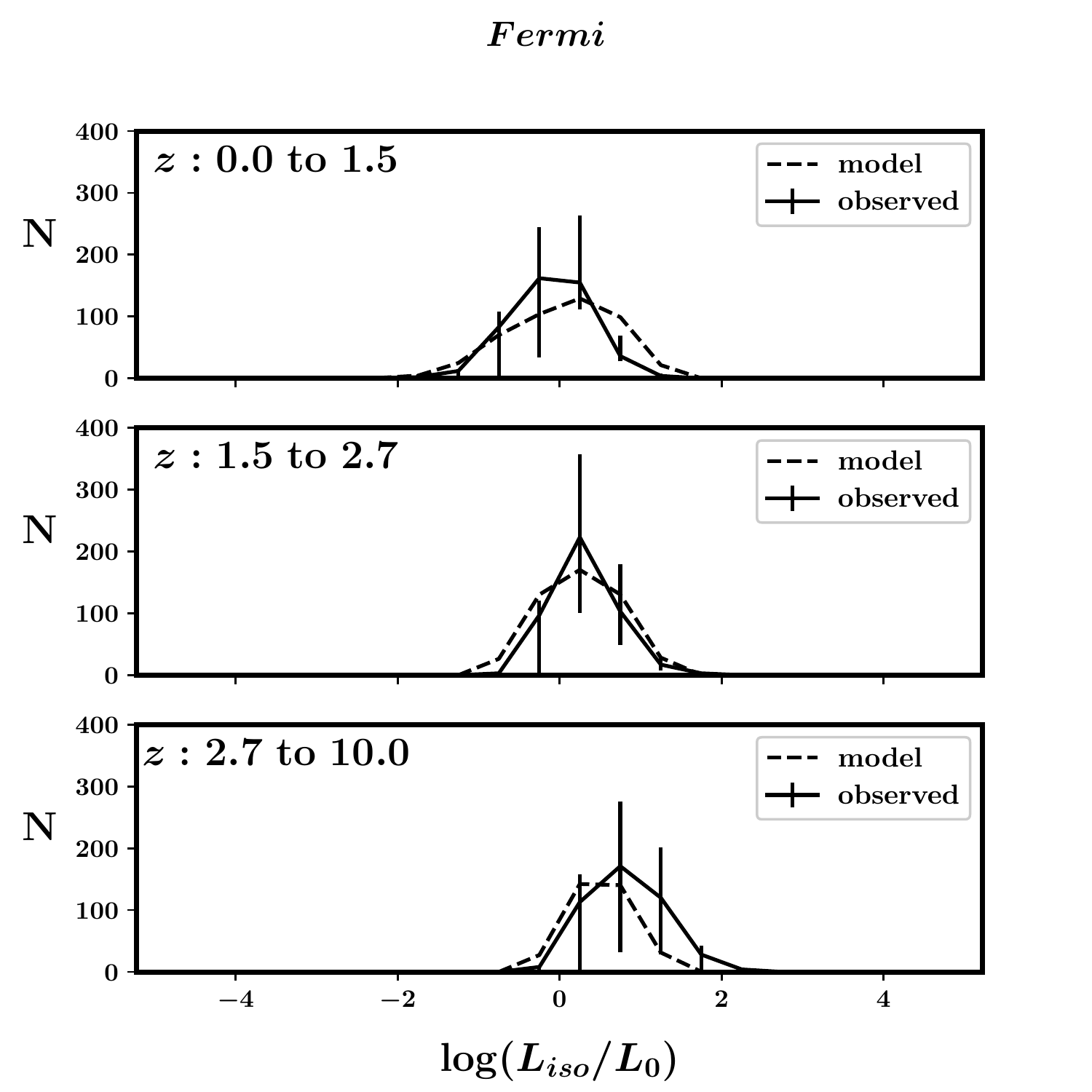}\includegraphics[scale=0.39]{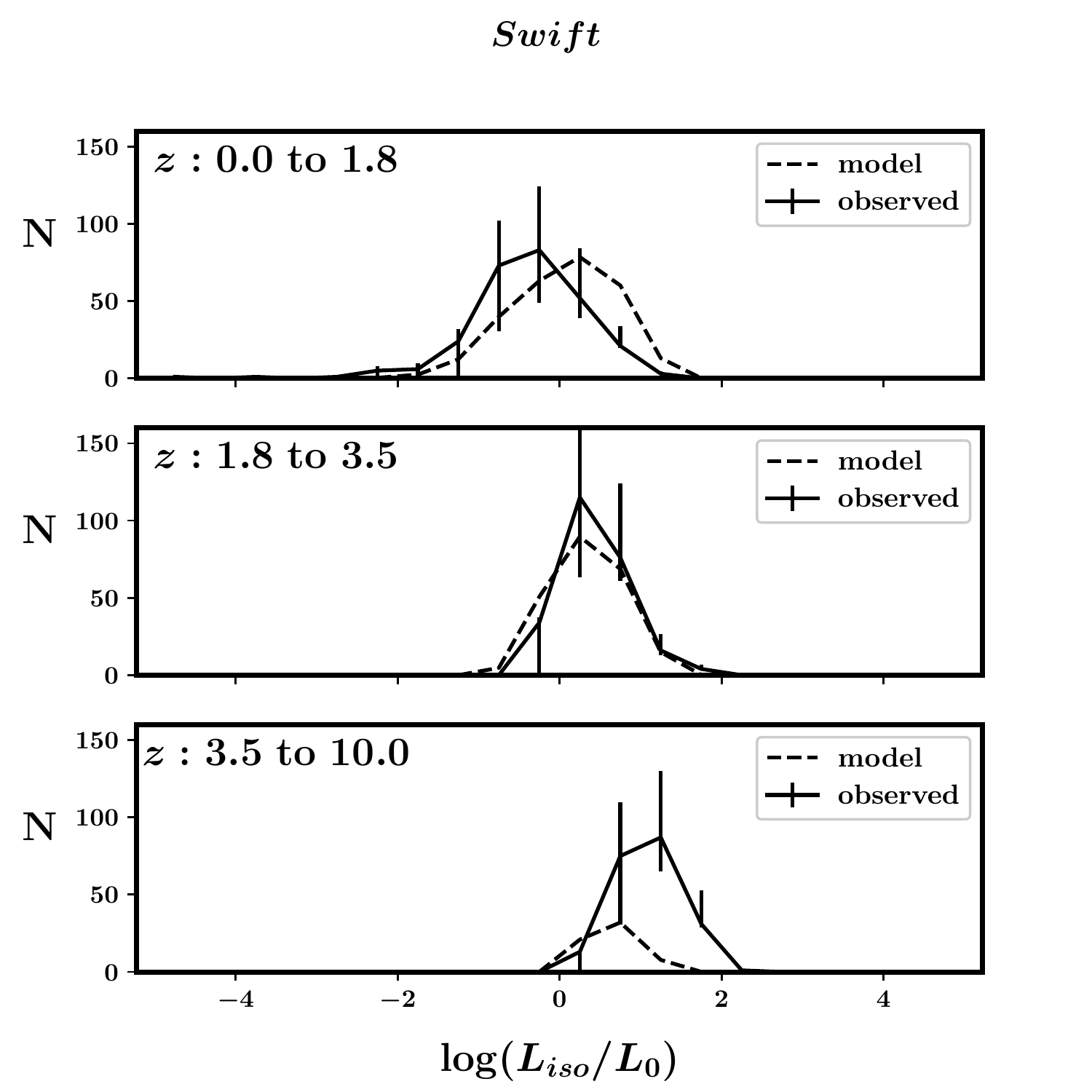}
\end{centering}
\begin{centering}
\includegraphics[scale=0.37]{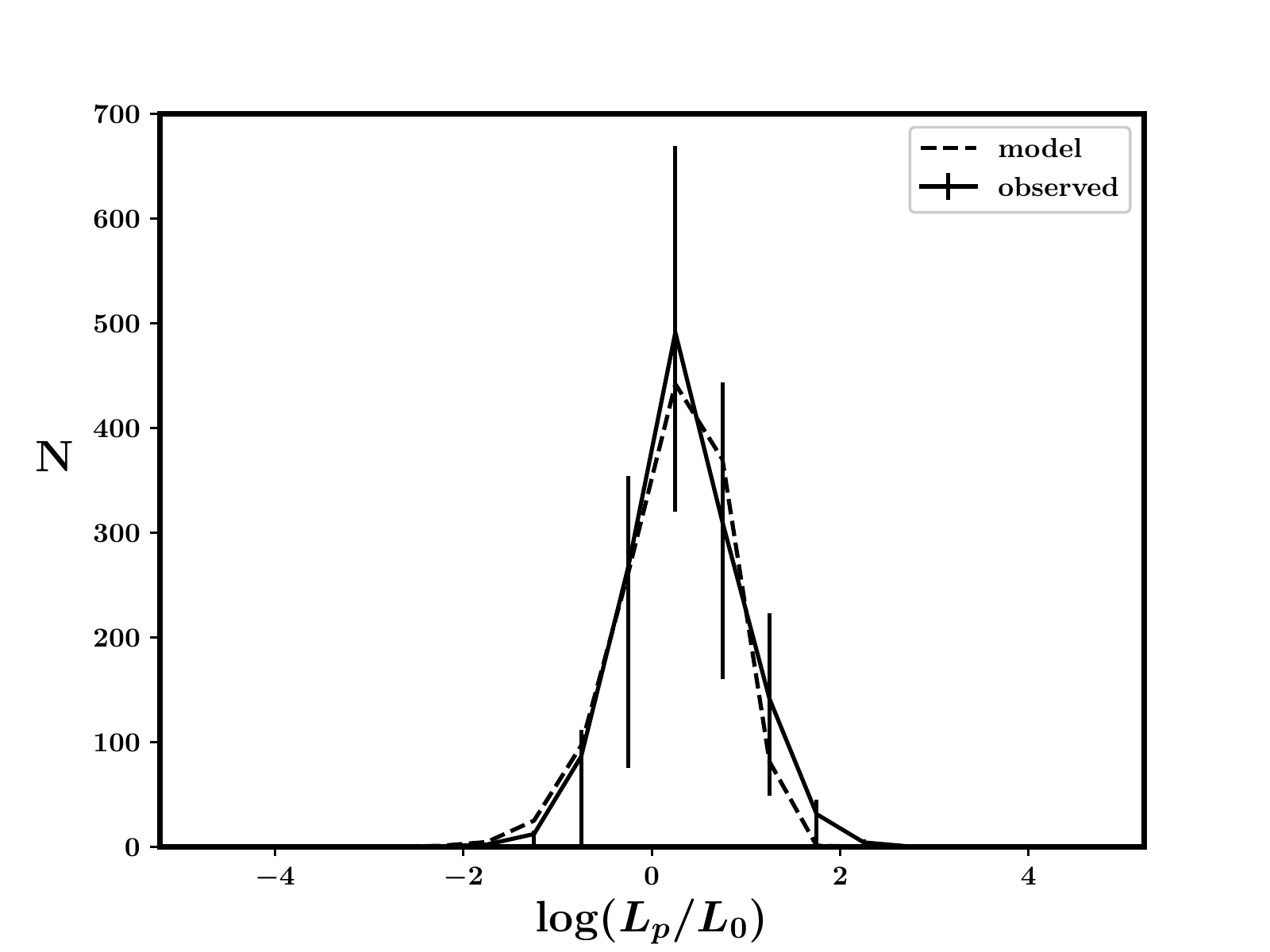}\includegraphics[scale=0.37]{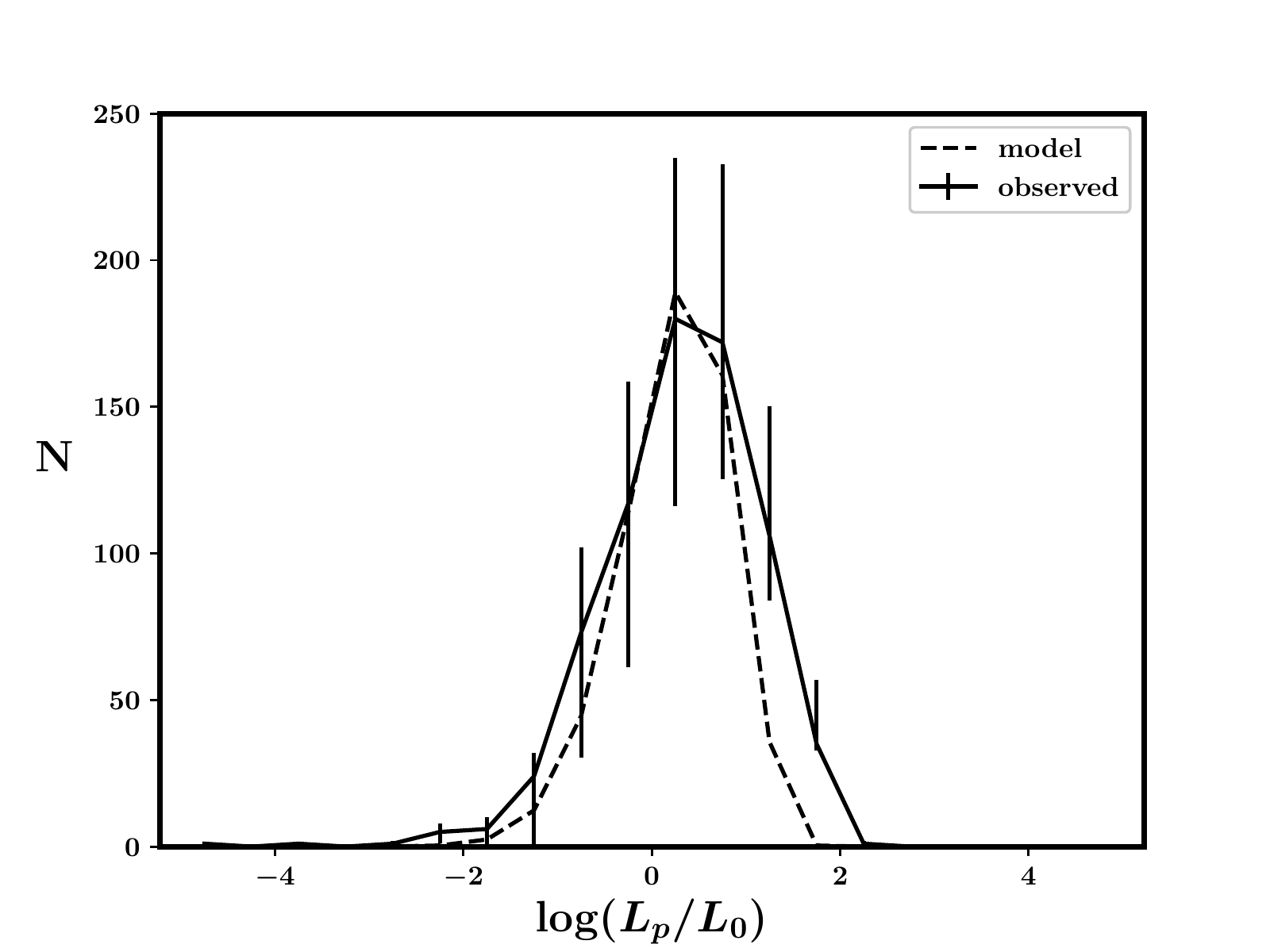}
\end{centering}
\centering{}\caption{The comparison of data and fits for the ECPL model. Upper panels: The observed distribution, binned according to equipopulous redshift bins, for each instrument. Lower panel: Integrated over redshift, for the corresponding instruments in the upper panel. Here, $L_{0}=10^{52}{\rm \, erg.s^{-1}.}$ The `model' refers to that described in the text, with the final solutions of the parameters tabulated in Table \ref{tab:ECPL_model_parameters}. The errors are derived by taking into account the derived errors on luminosities, while binning. Note that the discrepancies between the model and the data are quite large for the \s\, large redshift bins. This is due to the unaccounted detection probability, which is larger for \s\, at higher redshifts.\label{fig:bestfit_ECPL_models}}
\end{figure*}

\begin{figure*}
\begin{centering}
\includegraphics[scale=0.39]{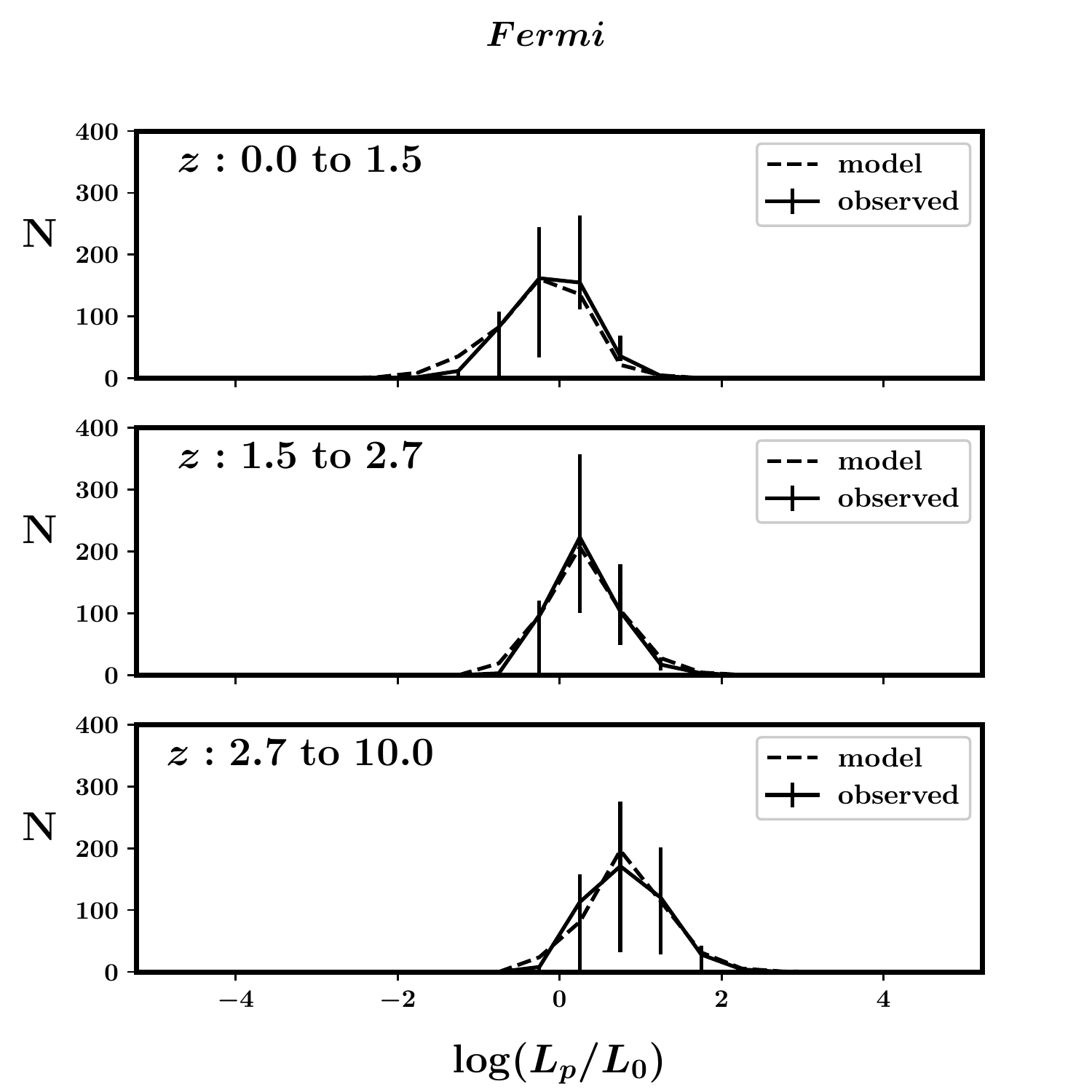}\includegraphics[scale=0.39]{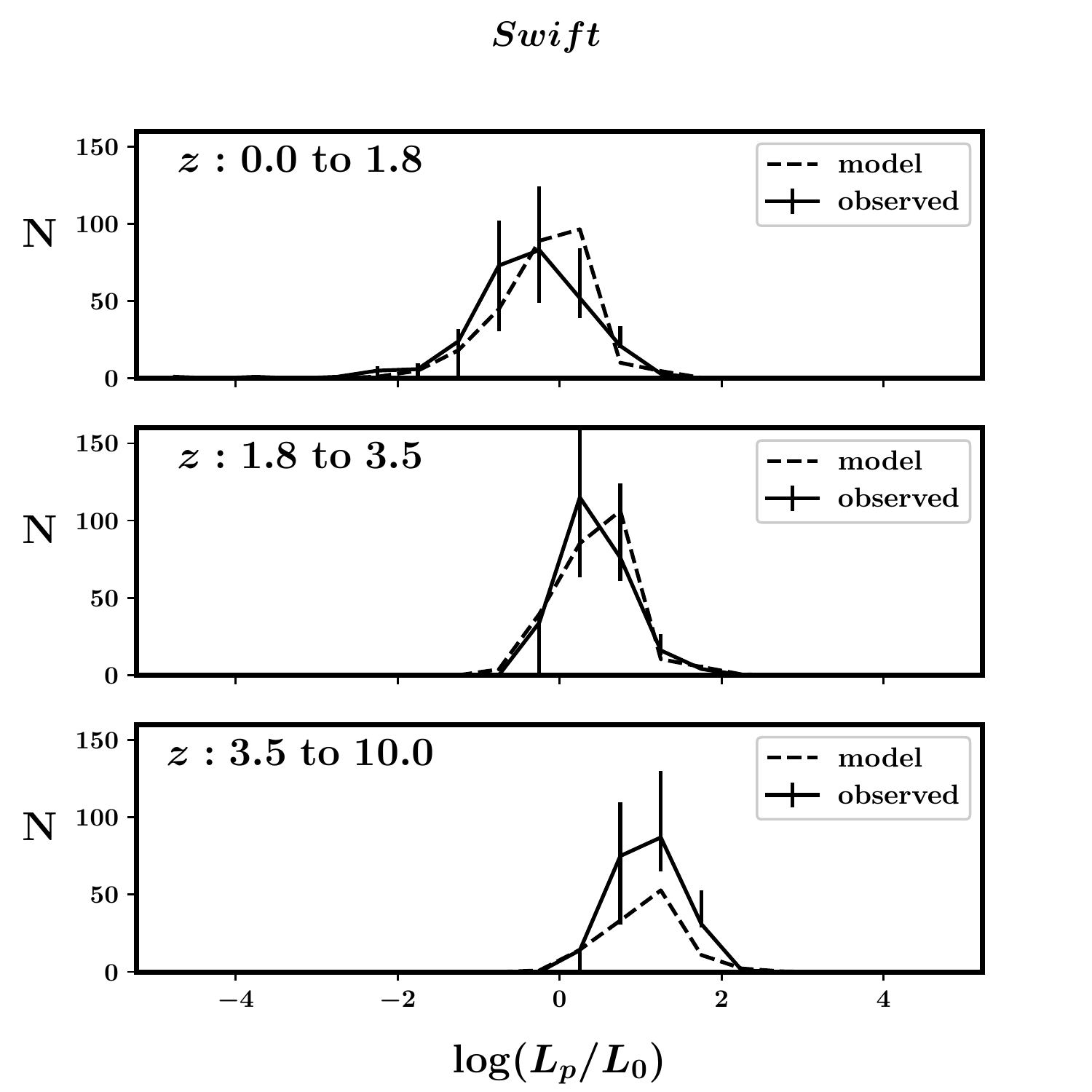}
\end{centering}
\begin{centering}
\includegraphics[scale=0.37]{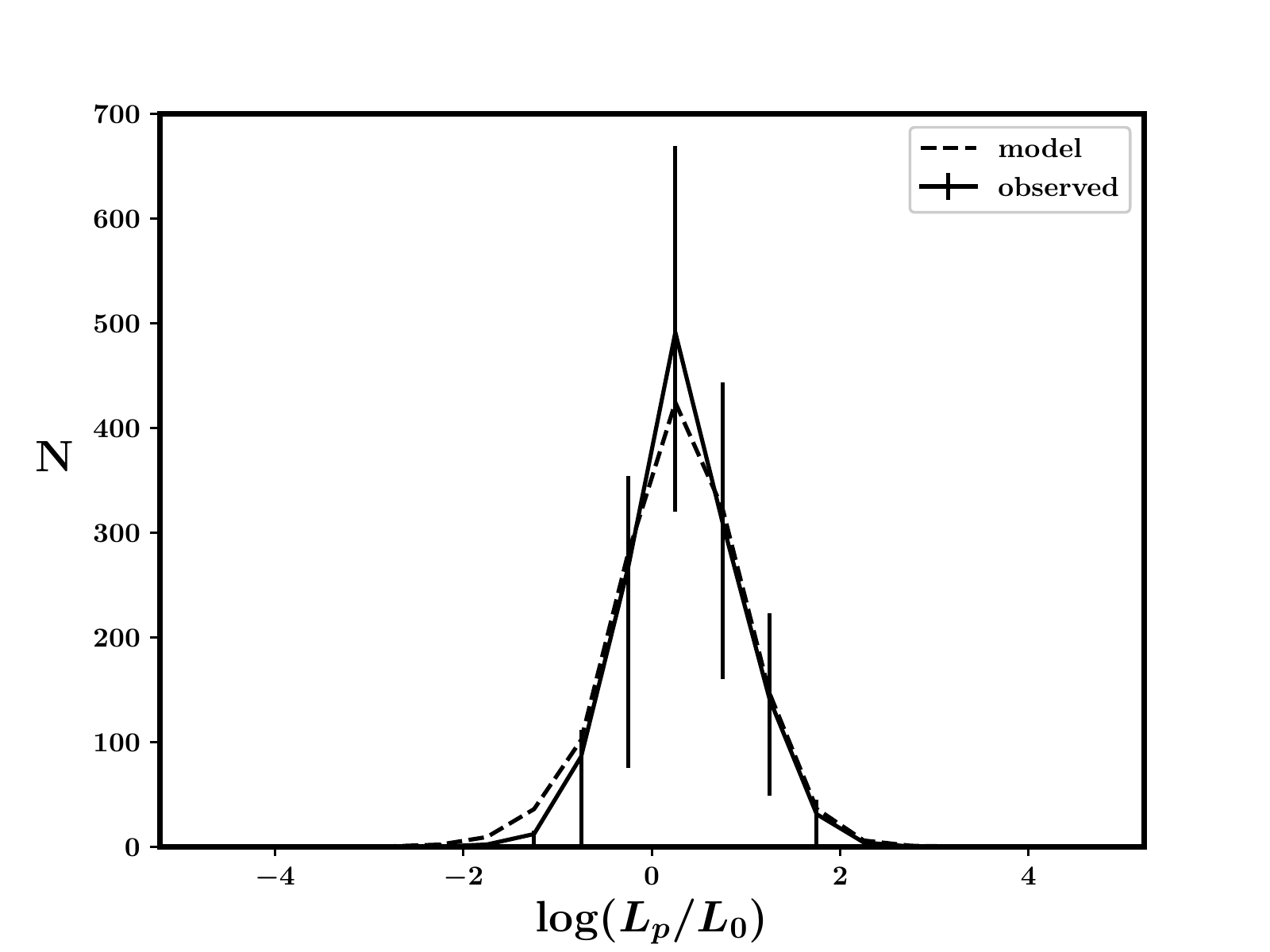}\includegraphics[scale=0.37]{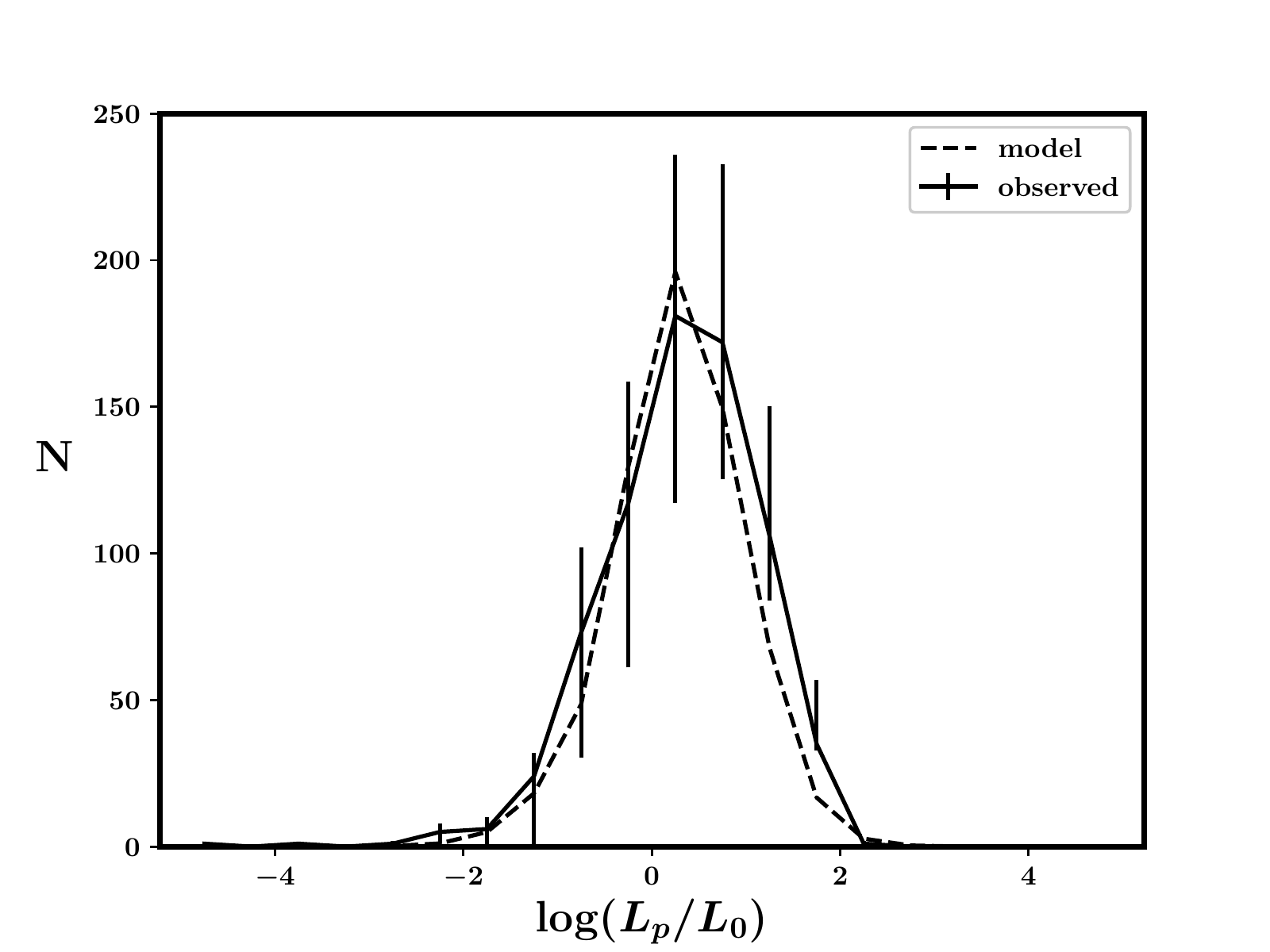}
\end{centering}
\centering{}\caption{Same as Fig \ref{fig:bestfit_ECPL_models}, for the BPL model. Parameters given in Table \ref{tab:BPL_model_parameters}.}\label{fig:bestfit_BPL_models}
\end{figure*}

I look for the best-fit parameters of each model for \f\, and \s\, GRBs separately, because they have different $L_{cut}\left(z\right)$ as shown in Fig \ref{fig:pseudo_redshifts_and_luminosities} (refer to Table \ref{tab:GRB_numbers} for the classes). 
For the case of the ECPL, it is noticed that any non-zero values of $\delta$ or $\chi$ (or both) decreases the quality of fit, for both \f\, and \s. This allows me to decrease the parameter-space into a 2-dimensional space of $\nu$ and $L_{b,0}$ (which is equal to $L_b$ for $\delta = 0.)$ In the case of the BPL however, the data strongly requires the inclusion of a positive-definite $\delta$ and a  negative-definite $\chi.$ It is to be noted that the ECPL has one parameter less than the BPL, but allows the break to vary naturally, explaining why the data requires the additional dependencies on the parameters $\delta$ and $\chi$ for the BPL model.

I search for the solutions by computing $d_{z}^{2}=\sum_{L}\left[N_{{\rm model}}\left(L,z\right)-N_{{\rm observed}}\left(L,z\right)\right]^{2}$ for each redshift bin, then evaluating the discrepancy $d^{2}=\sum_{z}d_{z}^{2},$ and finally looking for the model parameters that reduces $d^{2}.$ I optimize the search by first choosing a large grid of parameters with sufficiently small bins, and then gradually converge on the best-fit parameters by decreasing the search-space and bin-size at each run.

In the case of the ECPL, both the \f\, and \s\, runs converge to similar values of parameters, and are consistent within the deduced errors. The fits are generally poorer for the latter case, and also because \s\, detects a larger number of GRBs at higher redshifts due to its higher sensitivity compared to \f. This, however, is not directly taken into account in the modelling, being a limitation of the present work. This is because the exact mathematical form of the detection probabilities at various fluxes is not known. Hence, I tabulate the parameters from only the \f\, fits, in Table \ref{tab:ECPL_model_parameters}. The data is generally over-fitted, with the  the $\red$ for the two instruments being $0.116$ for \f\, and $0.539$ for \s. This is because of the large number of bursts with similar luminosities, all with similarly large uncertainties.

In the case of the BPL, there is no oscillation of any of the five parameters, justifying that the solutions are global. However it is found that \f\, and \s\, have different best-fits, significant differences being only in the related parameters $\nu_{2},$ $L_{b,0}$ and $\delta.$ The \s\, solutions require extreme evolution of the break luminosity $\left(\delta=3.95\right),$ and raises suspicion of being an artefact of unaccounted systematics. To understand this, I model the detection probabilities of the two instruments by a simple flux powerlaw model, and plugging in the retrieved parameters of the two instruments, find that the difference can be explained by the variation of the detection probabilities with redshift and luminosity. On further investigation, I find that the \s\, solutions are in fact degenerate with the \f\, solutions. The $d^{2}$ contours in the $L_{b,0}$-$\delta$ space have similar global shapes, and also behave similar locally around the \f\, solutions. Thus I conclude that the best-fit solutions obtained for \s\, are driven by complications of its detection probability, and hence choose the \f\, best-fits as the accepted solutions, thus breaking the degeneracy. These are tabulated in Table \ref{tab:BPL_model_parameters}. The corresponding fits for the two instruments are shown in Fig \ref{fig:bestfit_BPL_models}.  The larger proportional errors for \s\, make the $\red$ comparable for the two instruments however, $0.362$ for \f\, and $0.364$ for \s. This demonstrates that the use of \f\, bursts helps in solving the degeneracy of the parameter space of the model.

\begin{table*}
\caption{The best-fit parameters for the BPL model, as found by extensive search in the $5$-dimensional space. The convergence of the parameters are tested thoroughly. As a comparison, I show the best-fit parameters for the equivalent model of the recent works of \citet{Amaral-Rogers_et_al.-2017-MNRAS} and \citet{Tan_et_al.-2013-ApJL}. Since the GRB formation rate is modelled differently in the former, the parameter $\epsilon$ cannot be compared. Moreover, one needs to be cautious to expect the other parameters to agree for the same reason. However, except $\nu_{2},$ reasonable agreement is found. The comparison is straightforward with \citet{Tan_et_al.-2013-ApJL}, which however does not cite errors on their parameters. An overall agreement is noticed between the two works.
\label{tab:BPL_model_parameters}}
\begin{centering}
\begin{tabular}{|c|c|c|c|}
\hline 
parameter & present work & \citet{Amaral-Rogers_et_al.-2017-MNRAS} & \citet{Tan_et_al.-2013-ApJL}\tabularnewline
\hline 
\hline 
$\nu_{1}$ & $0.65_{-0.3}^{+0.1}$ & $0.69 \pm 0.09$ & $0.8$\tabularnewline
\hline 
$\nu_{2}$ & $3.10_{-0.4}^{+0.5}$ & $1.88 \pm 0.25$ & $2.0$\tabularnewline
\hline 
$L_{b,0}$ & $0.30_{-0.1}^{+0.15}$ & $0.15_{-0.09}^{+0.20}$ & $0.12$\tabularnewline
\hline 
$\delta$ & $2.90_{-0.50}^{+0.25}$ & $2.04 \pm 0.45$ & $2$\tabularnewline
\hline 
$\epsilon$ & $-0.80_{-1.0}^{+0.75}$ & - & $-1.0$\tabularnewline
\hline 
\end{tabular}
\par\end{centering}
\end{table*}

Since the constant in the RHS of Equation \ref{eq:fB.C_of_z} is not known a priori, it is calculated via the solutions of the models. It is known that for \f, $T\sim8.5$ yr and for \s, $T\sim12$ yr. I assume $\frac{\Delta\Omega}{4\pi}\sim\frac{1}{3}$ for \f\, and $\frac{1}{10}$ for \s, to get ratios of the observed and modelled numbers, which are converted to get

\begin{equation}
f_{B}C(0)=\begin{cases}
12.329\times10^{-8}\,{\rm M_{\odot}^{-1}}, & Fermi,\\
12.842\times10^{-8}\,{\rm M_{\odot}^{-1}}, & Swift.
\end{cases}\label{eq:fB.C_of_zero_ECPL}
\end{equation} for the ECPL model, and

\begin{equation}
f_{B}C(0)=\begin{cases}
7.498\times10^{-8}\,{\rm M_{\odot}^{-1}}, & Fermi,\\
8.200\times10^{-8}\,{\rm M_{\odot}^{-1}}, & Swift.
\end{cases}\label{eq:fB.C_of_zero_BPL}
\end{equation} for the BPL model.

\begin{flushleft}
These numbers are consistent with each other, and in rough agreement with those quoted by \citet{Tan_et_al.-2013-ApJL}.
\par\end{flushleft}

The ECPL shows agreement with the most recent work of \citet{Amaral-Rogers_et_al.-2017-MNRAS}. The BPL model shows a sharp change at its break, which itself evolves quite strongly with redshift as $L_{b}\sim0.3\times10^{52}\left(1+z\right)^{2.90} \, \rm{erg.s^{-1}},$ in general agreement with \citet{Amaral-Rogers_et_al.-2017-MNRAS}. The GRB formation rate for a given star-formation rate decreases with increasing redshift as $f_{B}C \propto (1+z)^{-0.80}$ (the normalization is given by Equation \ref{eq:fB.C_of_zero_BPL}), in agreement to the reports of \citet{Tan_et_al.-2013-ApJL}. Whereas the ECPL automatically takes into account the variation of the break, this needs to be incorporated via  strong evolutions with redshift in the BPL model. However, it is not possible to distinguish between the two models based on the fits. One of the reasons is that the data is generally over-fitted due to the large uncertainties, and another possible reason being that the discrepancies between data and model could be a result of the complex nature of detection probabilities of the instruments, which I have not attempted to model directly.

It is to be noted that the present work is empirical; it does not attempt to provide an understanding of the models used, nor of the derived values of the parameters. A thorough understanding of the observed GRB number distribution requires one to justify the models via the phenomenology of long GRBs, taking into consideration the beaming of GRB jets and the GRB formation environment. This the scope of future work.

\subsection*{Predictions for CZTI}

The CZT Imager or CZTI \citep{Bhalerao_et_al.-2016-arXiv-JAA}, on the Indian multi-wavelength observatory $\AS$ \citep{Rao_et_al.-2016-arXiv-Astrosat} is capable of detecting transients at wide off-axis angles, localizing them to a few degrees, and carrying out spectroscopic and polarization studies of GRBs, as demonstrated in \citet{Rao_et_al.-2016-ApJ}. A preliminary analysis done with the weakest GRB detected by CZTI suggests that it is at least as sensitive as \f, which detects roughly $3$ times the number of GRBs per year compared to \s. Similar to \f, the CZTI is also a wide-field detector. Moreover, it covers a wide energy range, being the most sensitive between $50$ and $200$ keV. Thus, it is reasonable to assume that its GRB detection rate is at least comparable to that of \s. Assuming this, I make predictions for CZTI over the redshift bins that were chosen for \f. The best-fit ECPL model predicts that CZTI should detect $150$ GRBs per year. The best-fit BPL model predicts detection-rate of around $140$ GRBs per year, with the \f\, equipopulous redshift bins almost equipopulous for CZTI as well. In $\sim1.3$ years of operation, $\sim 120$ GRBs has been detected by CZTI by triggered searches alone,\footnote{See a comprehensive list at \href{http://astrosat.iucaa.in/czti/?q=grb}{http://astrosat.iucaa.in/czti/?q=grb}.} however the exact number is subjective. An automated algorithm to detect GRBs is being thoroughly tested and implemented, the details of which will be reported elsewhere. In the view of this, the predictions point out the fact $\sim 20$-$30$ GRBs are yet undiscovered from the CZTI data. This is encouraging for the efforts on automatic detection, as well as that of the quick localization and follow-up, which will also be reported elsewhere.

\section{Conclusions}
\label{sec:Conclusions}

Previously, BATSE and \s\, GRBs have been used to constrain the GRB luminosity function. Only a few BATSE GRBs had redshift measurements, so indirect methods were used to study the luminosity function of these GRBs. On the other hand, about $30\%$ of the \s\, GRBs have redshift measurements. However, the measurement of the spectral parameters are also crucial to the measurement of the luminosity, via the $k$-correction factor. Being limited in the energy coverage, estimates of the \s\, spectral parameters have large uncertainties. Moreover, the number of \s\, GRBs with redshift measures are not as large as the entire BATSE sample. \f\, is a GRB detector with large sky coverage, a detection rate roughly $3$ times more than \s, and wide energy coverage, thus measuring the broad-band spectrum of a large fraction ($\sim75\%$) of the detected GRBs to sufficient accuracy. However, its poor localization capabilities makes it impossible to make \s-like follow up observations, and hence the measurement of redshifts.

In this work, I show that one of the methods proposed to solve the absence of redshift measures for BATSE GRBs can be used self-consistently to estimate the luminosities of \s\, and \f\, GRBs without redshift measurements. This method works on the premise that the `Yonetoku correlation' is applicable to all GRBs. For this, I have first used the most updated common sample of $66$ long GRBs detected by these two instruments, to re-derive the parameters of this correlation. By a careful study of the discrepancies, I find a significant trend between the ratio of the observed and predicted luminosities with the measured redshift. The exact reason for this trend is not clear, but it highlights the fact that the weakness of the correlation is intrinsic, being driven by physical effects and not measurement uncertainties. I conclude that although the large scatter in the Yonetoku correlation rules out the possibility of using GRBs as distance-indicators, the statistical distribution of observed redshifts is reproduced well, and there is no need to modify the extraction of the correlation parameters as has been suggested previously \citep{Tan_et_al.-2013-ApJL}.

Next, the method is shown to self-consistently predict `pseudo redshifts' of all long GRBs without redshift measurements. This allows calculation of the luminosities of a total of $2067$ GRBs from these instruments, including the subsample (of $66$ bursts) that has direct measurements of both redshift and spectra. I then use this large sample to model the GRB luminosity function, and place constraints on two models. The GRB formation rate is assumed to be a product of the cosmic star formation rate and a GRB formation efficiency for a given stellar mass. Whereas an exponential cut-off powerlaw model does not require a cosmological evolution, a broken powerlaw model requires strong cosmological evolution of both the break as well as the GRB formation efficiency (degenerate upto the beaming factor of GRBs). This is the first time \f\, GRBs have been used independent of measured redshifts from \s\, to study the long GRB luminosity function. Moreover, this is the first time such a large sample of \s\, GRBs have been used. The use of the large sample of \f\, GRBs helps in placing sufficient confidence on the derived parameters of the broken powerlaw model, when \s\, GRBs alone suffer from degeneracies and observational biases. Comparison with recent studies shows reasonable agreement for both the models, however it is not possible to distinguish between them.

\citet{Amaral-Rogers_et_al.-2017-MNRAS} has proposed on increasing the sample of GRBs by taking individual pulses of the same bursts as physically separate entities. In the future, perhaps a conglomeration of their method with the one here can be implemented to increase the sample size even further, to further test the parameters of the models and also carry out an in-depth analysis of the detection probabilities of the two instruments, which is presently quite a daunting task. This work also does not attempt to provide a physical understanding of the empirical models or the parameter values derived, which should be addressed in future works.

Finally, I have used the derived models as templates to make predictions about the detection rate of GRBs by CZTI on board $\AS.$ The predictions are encouraging for the ongoing efforts of this collaboration. The quick localization of the few bursts that are predicted to be detected only by CZTI can increase the GRB database even further, and reveal interesting answers about the GRB phenomenon in both the local and the distant universe.

\section*{Acknowledgements}
\label{sec:Acknowledgements}
I sincerely thank my Ph.D. advisor A. R. Rao for helpful discussions and suggestions during the entire course of the work; Pawan Kumar for his comments on the manuscript, thus improving its quality; and the referee for his critical comments which immensely improved the quality of the work.

\end{document}